# An *ab initio* molecular dynamics exploration of associates in Ba-Bi liquid with strong ordering trends

Jianbo Ma,[1,2] Shun-Li Shang,[1] Hojong Kim,[1] and Zi-Kui Liu[1]

[1]Department of Materials Science and Engineering, The Pennsylvania State University, University Park, PA, 16802, USA

[2]School of Material Science and Engineering, Shanghai Jiao Tong University, Shanghai 200240, China

**Abstracts:**

Fictive associates are widely used to describe and model liquid phases with strong ordering trends. However, little evidence is known about the assumed associates in most cases. In the present work, an *ab initio* molecular dynamics (AIMD) study is employed to investigate the characters of the Ba-Bi liquid, in which associates have been assumed in existing thermodynamic modeling. It is found that in the Ba rich melt, the Bi atoms are almost completely surrounded by Ba atoms. The Bi-centered coordination polyhedrons are strongly associated to crystalline structures of $Ba_5Bi_3$ and $Ba_4Bi_3$ with a longer lifetime than other polyhedrons during the AIMD simulations. In addition, these Bi-centered polyhedrons in Ba rich melt connect with each other through vertex, edge, face, and/or bipyramid sharing to form medium range orders (MRO). In the Bi rich melt, the Ba-centered polyhedrons also form MROs, but they are both structurally and compositionally diverse with a shorter lifetime. These findings from AIMD study provide evidences that there exist a strongly ordering $Ba_4Bi_3$ associate and a weakly ordering $BaBi_3$ associate in the Ba-Bi liquid. The predicted enthalpy of mixing in the liquid agrees well with the results by the CALPHAD modeling in the literature.

# 1.Introduction

Recently, Lichtenstein et al. [1] reported the extremely low chemical activity of Ba ($a_{Ba}$) in the Bi rich melt by measuring the electromotive force (EMF) of the Ba-Bi alloys. For example, at the barium mole fraction $x_{Ba} = 0.05$, the $a_{Ba}$ value is as low as $6.3 \times 10^{-16}$ at temperature $T = 773$ K; and $a_{Ba} = 8.7 \times 10^{-12}$ at $T = 1123$ K. These EMF results indicate that there exist strong chemical affinity and short-range order between Ba and Bi, i.e., there are chemical associates [2-4] in the liquid.

There are two major thermodynamic approaches to account for strong short-range ordering, i.e., (i) the quasichemical model using pair approximation [5] and (ii) the associate model assuming a mixture of liquid single element species and molecule-like associates [3, 4]. In the associate model, the compositions of the associates are usually the same as those of the intermetallic compounds with high melting temperatures or highly negative values of enthalpy of mixing [3, 4]. The associate model has been successfully applied to model thermodynamic properties of binary and multi-component systems [2, 3, 6-12].

Atomic structures in molten or amorphous alloys are usually considered to be ordered in short up to medium range with the existence of hexagonal, pentagonal, and/or quadrangular bipyramid entities, such as in the molten Al [13], Ni [14], Zr [15], glass $Zr_{80}Pt_{20}$ [16], glass $Al_{75}Ni_{25}$ [16], and liquid/glass Cu-Zr [17]. Solute-solvent atomic network (e. g., in glass $Ni_{63}Nb_{37}$ [16] and liquid $Ni_3Al$ [18]) or connected pentagonal bi-pyramids (e.g., in molten $Al_3Cu$ [19]) can also form medium range orders (MRO). It has been observed that the structures of solute centered polyhedrons depend strongly on chemistry and the ratios of their atomic radii [16]. It is believed that compacted polyhedrons can retard the crystallization process [20-22], resulting in supercooled melt, amorphous alloy, or segregation of alloying elements [23].

In a recent CALPHAD (calculation of phase diagram) modeling [24] of the Ba-Bi system, Liu et al. [10] used two fictive associates of $Ba_4Bi_3$ and $BaBi_3$ in the liquid phase, which were empirically selected based on the intermetallic compounds with the two highest congruent melting points. The present investigation aims to use the *ab initio* molecular dynamics (AIMD) simulations to study the presence and the characters of associates in the Ba-Bi liquid phase. The AIMD approach [25] uses the interatomic interactions calculated on the fly based on density functional theory (DFT), avoiding the errors due to empirical potentials.

## 2. Calculation details and structure analysis methods
### 2.1 AIMD calculations

The AIMD simulations for molten $Ba_{1-x}Bi_x$ (where $x$ is the mole fraction of Bi species) were carried out using Vienna *Ab-initio* Simulation Package (VASP) [26-28] with the exchange-correlation functional described by the generalized gradient approximation (GGA) parameterized by Perdew and Wang [29, 30]. The electronic configurations of $5s^25p^66s^2$ and $6s^26p^3$ were employed as valences for Ba and Bi, respectively. Cubic supercells containing 200 atoms with periodic boundary and the NVT canonical ensemble (i.e., the fixed number of atoms, volume, and temperature), were employed for these AIMD simulations. The simulation temperature for each supercell was 1123 K, which is above the temperatures of liquidus in the Ba-Bi system and is also one of the temperatures in the previous EMF measurements [1]. The plane-wave cutoff energy was set as 200 eV, which is above the default energies suggested by VASP, i.e., 187.2 eV for Ba and 105.0 eV for Bi. Using this high cutoff energy, the Pulay stress can be ignored [31]. The Brillouin zone was sampled at a single Γ-point. All the AIMD simulations were run for 10,000 steps with a time step of 3 femtoseconds. The first 4,000 steps were used to reach kinetic equilibrium of the AIMD simulations, and the configurations of the rest 6,000 steps were employed to analyze the structure and thermodynamic properties.

In the present AIMD simulations, the volumes of the supercell sizes as a function of composition were adjusted carefully to ensure the total external pressure (contributed by two parts, i.e., the movement of the atoms and the interaction among the atoms [32]) being close to zero. As an example, the cell length of the cubic supercell and the total external pressures for the molten $Ba_{1-x}Bi_x$ are plotted in the insets of

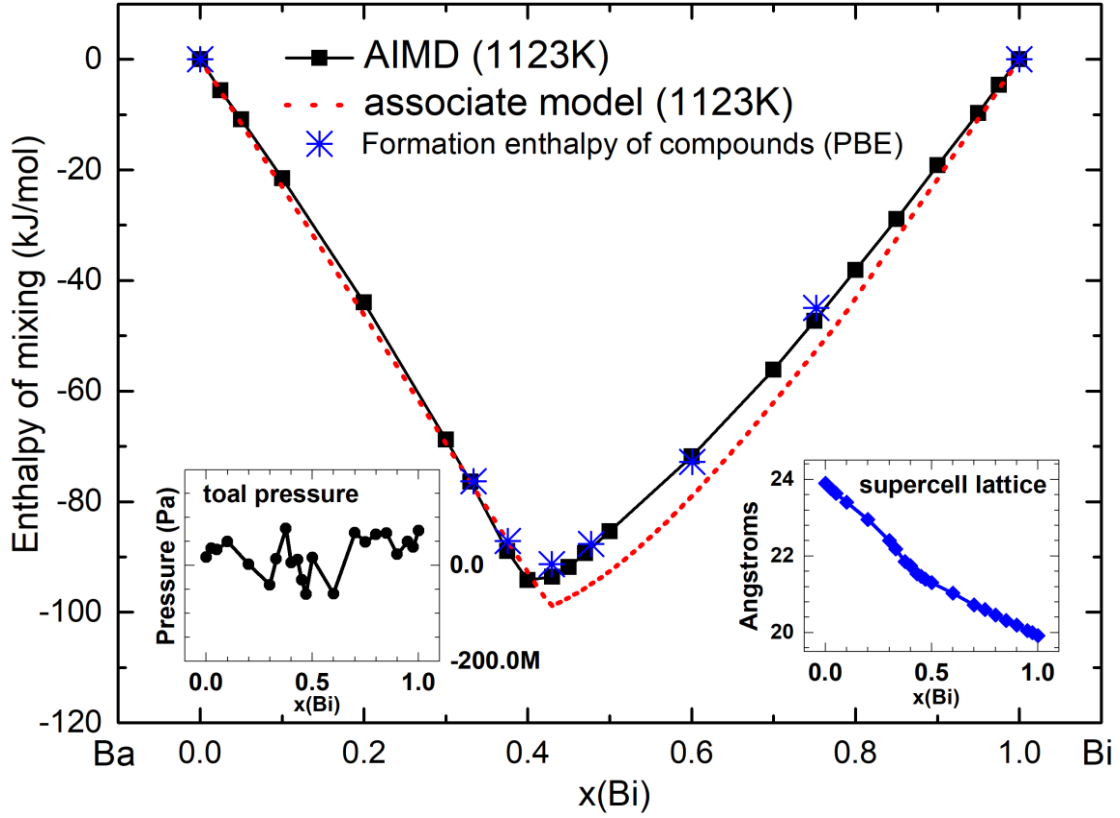

**Figure 1**, showing the decrease of cell length with increasing Bi content and the total external pressure lower than 0.1 GPa. It is worth mentioning that the atomic density of molten Ba determined in the present work, 0.0147 Å$^{-3}$ at 1123 K, is in good agreement with the experimental value of 0.0146 Å$^{-3}$ at 1001 K [33]. The density of molten Bi calculated in the present work, 0.0253 Å$^{-3}$ at 1123 K, is only slightly lower than the experimental value of 0.0266 Å$^{-3}$ at the same temperature [34].

The enthalpy of mixing ($\Delta H_{mix}$) of molten Ba$_{1-x}$Bi$_x$ at 1123 K (illustrated in Figure 1) is calculated by the formula (1)

$$\Delta H_{mix} = H_{mixture} - \sum c_i H_i \qquad (1)$$

where $H_{mixture}$ is the enthalpy of the system represented by the mean structural energy, $\langle E_0 \rangle$, obtained from AIMD, $c_i$ the mole fraction of component $i$, and $H_i$ the enthalpy of pure component Ba or Bi.

The details of structural analysis methods employed in present work are described as follows, from Sec 2.2 to 2.5.

**2.2 Pair correlation functions**

The partial pair correlation functions (PCFs) is calculated in terms of the following equation [31],

$$g_{\alpha\beta}(r) = \frac{L^3}{N_\alpha N_\beta} \left\langle \sum_{i=1}^{N_\alpha} \frac{n_{i\beta}(r)}{4\pi r^2 \Delta r} \right\rangle \quad (2)$$

where $L$ is the supercell length, $N_\alpha$ and $N_\beta$ are the total numbers of atoms for species $\alpha$ and $\beta$, respectively. $n_{i\beta}(r)$ is the number of atoms in the shell from distance $r$ to $r+\Delta r$ for species $\beta$ with atom $i$ of species $\alpha$ at the center. The angle brackets denote the time average of the system.

The first sharp peak on the PCF curves corresponds to the nearest neighbors of the central atom, and its location is roughly the mean bond length or the size of atom.

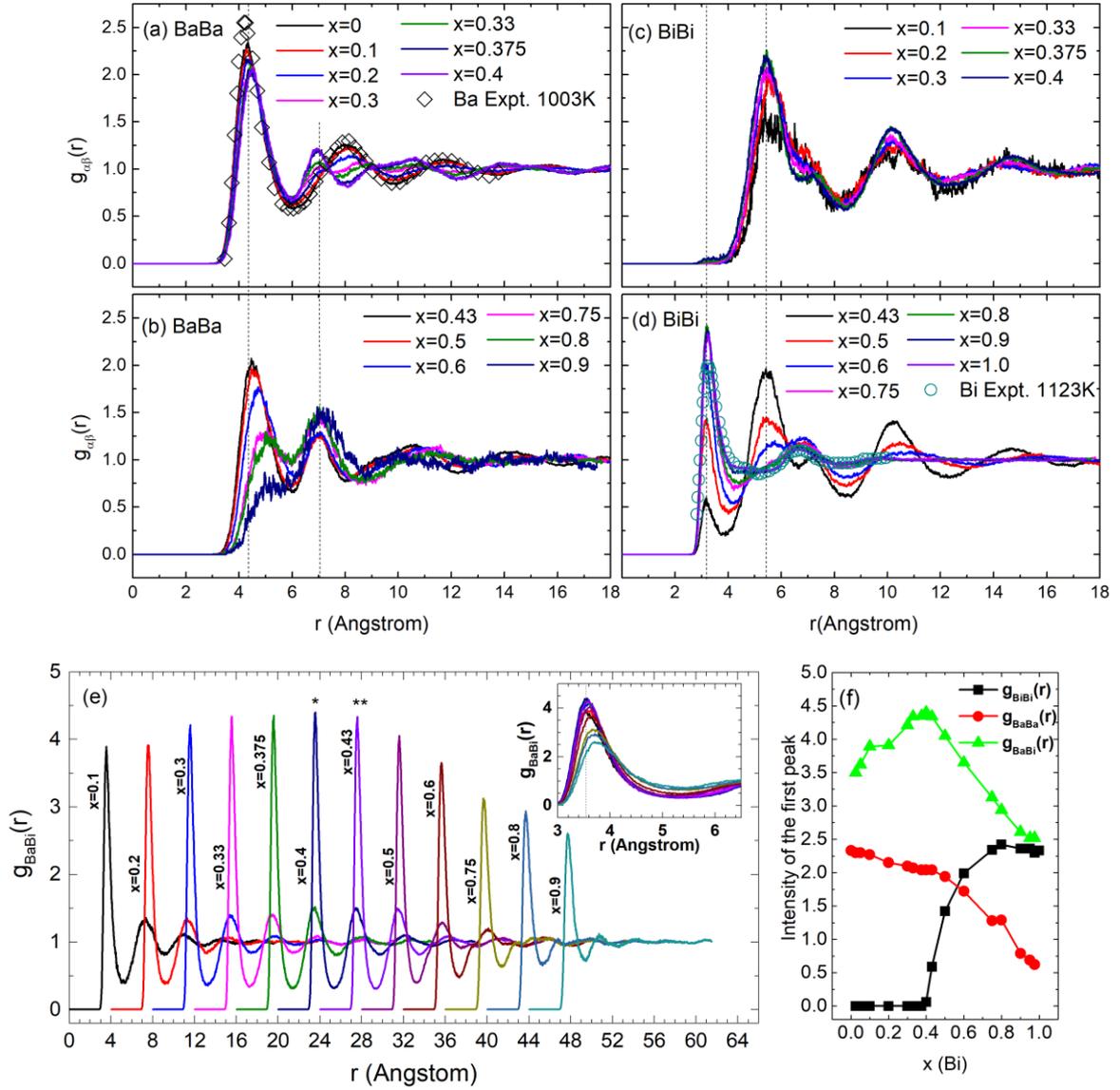

**Figure 2**a shows that the first and the second peaks on the PCF curves of molten Ba locate at 4.30 ± 0.02 Å and 8.01 ± 0.02 Å, respectively, which are in good agreement with the results measured by X-ray diffraction, i.e., 4.2 Å and 8.0 Å at 1003 K [35], respectively. Similarly,

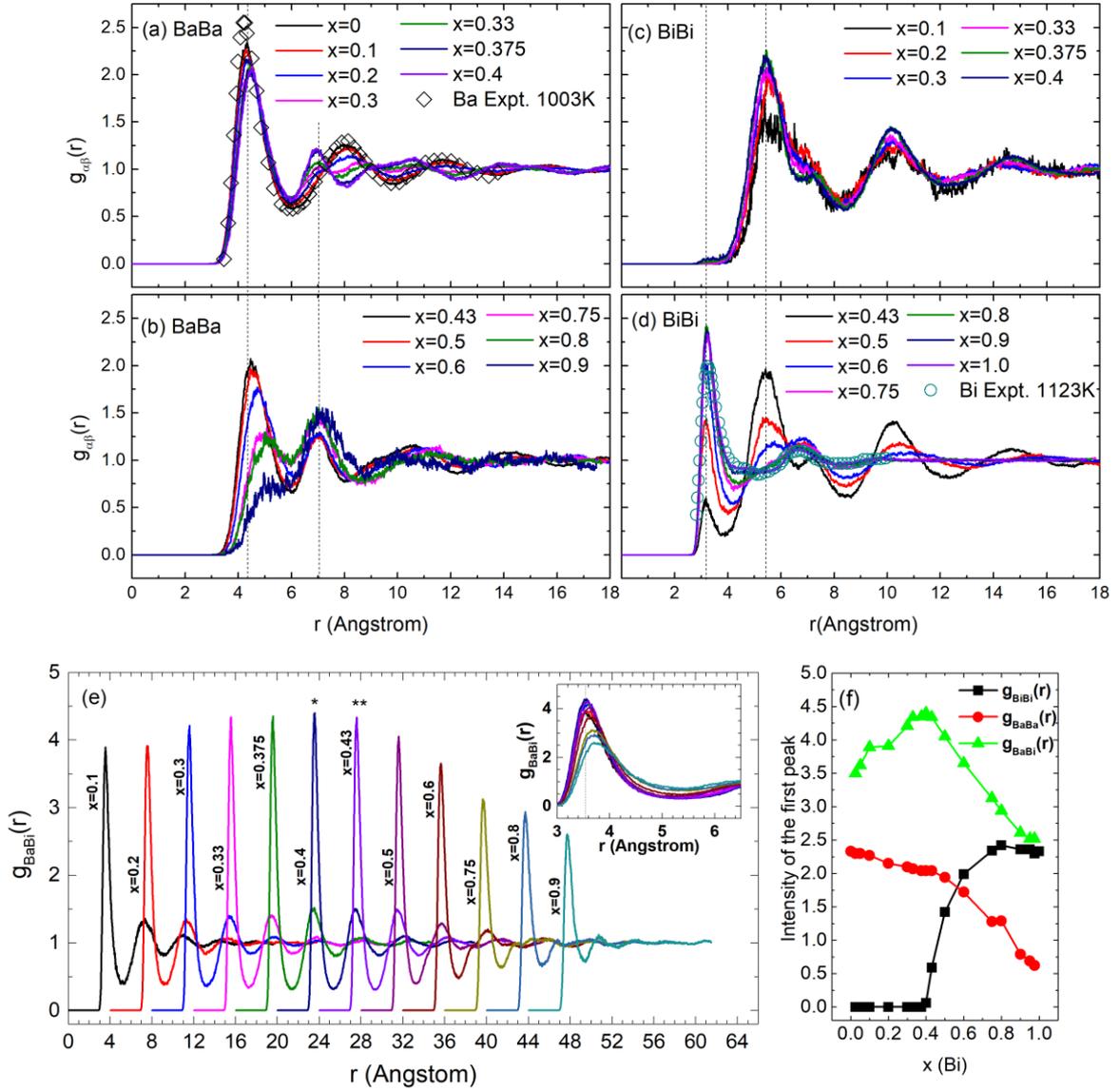

**Figure 2**d shows that the predicted first sharp peak location of molten Bi (3.25 ± 0.02 Å) is in excellent agreement with the measured value by neutron scattering, 3.25 Å at 1123 K [34], and other AIMD predictions, e.g., 3.25 Å at 1023 K [36] and 3.30 Å at 600 K [37].

A deep valley separates the first and the second peaks, and its location can be employed as a cutoff distance of bonding, $r_{cut}$. When the distance of atomic pair is shorter than $r_{cut}$, it is bonded and *vice versa*. The $r_{cut}$ values for the Ba-Bi pairs are 5.2, 5.3, and 5.4 Å when the $x$ values are in the ranges of 0-0.2, 0.3-0.375, and 0.4-1.0, respectively. A fixed value of $r_{cut}$ = 6.0 Å is employed herein for the Ba-Ba pairs. A fixed value of $r_{cut}$ = 4.2 Å is employed for the Bi-Bi pair. Note that these cutoff distances are also used to calculate the coordination number (CN).

The partial CNs (shown in **Error! Reference source not found.**) were calculated as follows,

$$Z_{\alpha\beta} = 4\pi\rho_\beta \int_0^{r_{cut}} r^2 g_{\alpha\beta}(r)dr \tag{3}$$

where $\rho_\beta$ is atomic density of species $\beta$. The Bi- and Ba-centered CNs, $Z_{Bi}$ and $Z_{Ba}$, are calculated by $Z_{Bi} = Z_{BiBi} + Z_{BiBa}$ and $Z_{Ba} = Z_{BaBa} + Z_{BaBi}$, respectively.

## 2.3 Static structure factor

The partial static structure factor (SF), shown in

$$S_{\alpha\beta}(q) = 1 + 4\pi\sqrt{\rho_\alpha\rho_\beta} \int [g_{\alpha\beta}(r) - 1]\frac{\sin qr}{qr} r^2 dr, \tag{4}$$

where $q$ is the wave vector in the reciprocal space, and $\rho_\alpha$ and $\rho_\beta$ are atomic densities for species of $α$ and $β$, respectively.

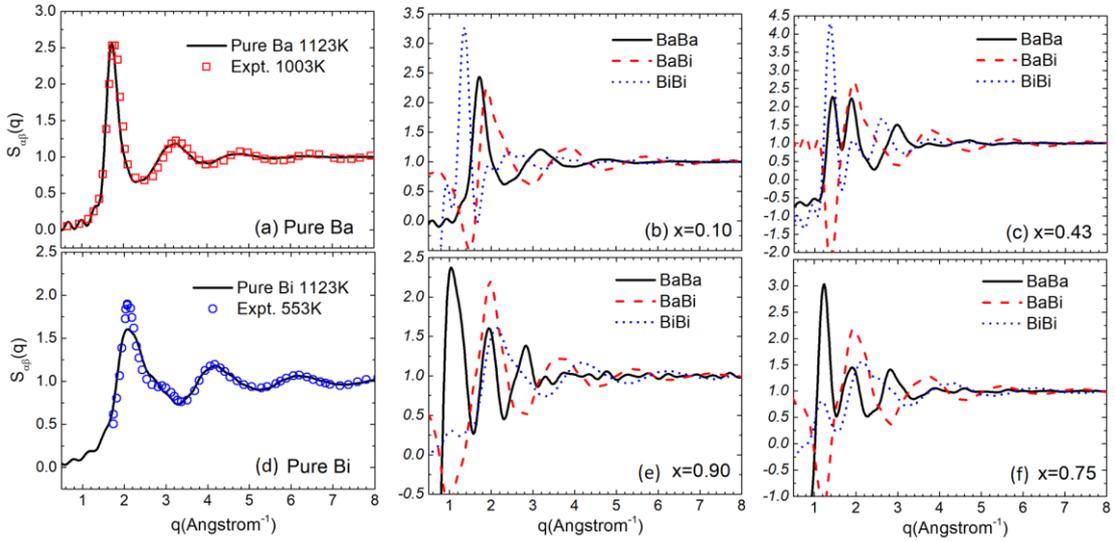

Figure 4, were obtained by Fourier transform from the partial PCF [38],

The calculated SFs of Ba and Bi are in good agreement with the experimental results [35, 39]; see Figure 4a and d. The locations of the main SF peaks of Ba and Bi, near q ≈ 1.73 Å$^{-1}$ and 2.09 Å$^{-1}$; respectively, match well the experimental results for molten Ba at 1003 K [35] and Bi at 553 K [39].

**2.4 Honeycutt-Andersen analysis**

To characterize the three-dimensional (3D) structures of the bonded atomic pair with their common neighbors, the Honeycutt-Andersen (H-A) analysis [40], also known as the common neighbor analysis [41], was performed. The H-A indices [40] with four integers, "*ijkl*", are employed to distinguish different atomic pairs. The first integer is to distinguish the bonding of atomic pair ($i = 1$) or not ($i = 2$). The second integer "*j*" is for the total common neighbors of the bonded root atomic pair. The third index, *k*, denotes the total bonds among these common neighbors. The last one is to distinguish the bonded-pair with the same first three indices, but different structures. For example, the bonded-pairs with indices 1661, 1551, 1441, and 1331, together with their common neighbors, are for structures of hexagonal, pentagonal, tetragonal, and trigonal bipyramid, respectively. When one of the bonds among these common neighbors is broken, they become 1651, 1541, 1431, and 1321, respectively. The bonded-pair with indices 1431 also can be transformed from 1551 with one of the common neighbors escaped. When one of the bonds among the common neighbors of 1431 is broken, it transforms to 1421 or 1422. The bonded-pair with indices 1001 have no common neighbor, such as the Bi-Bi atomic pair in crystalline Bi with rhombohedral A7 structure (Rhom_A7); see Table 1.

The fractions of the Ba-Ba, Ba-Bi, and Bi-Bi rooted bonded-pairs with different H-A indices were calculated (shown in Figure 6-8). The fraction is defined as the total bonded-pairs with the same indices divided by the total rooted bonds. As a comparison, the results of the common neighbor analyses for crystalline Ba, $Ba_2Bi$, $Ba_5Bi_3$, $Ba_4Bi_3$, $Ba_{11}Bi_{10}$, $BaBi_3$, and Bi are listed in Table 1.

## 2.5 Characterization of coordination polyhedron

For the purpose to distinguish coordination polyhedrons, a set of index, defined as $(l_2,l_3,…l_j)$ [13, 18], is employed, as inspired from the definition of Voronoi index [42]. Index "$l_j$" denotes the total number of the $ljxx$-type bonded-pairs around the central atom. The sum of "$l_j$", $\sum l_j$, is the CN. This method provides more details of coordination polyhedrons than the Voronoi index.

The Ba-centered coordination polyhedron (BAP) in BCC Ba is indexed as $(6_4,8_6)$ since each Ba atom is surrounded by six 1441 and eight 1661. Each Bi atom in crystalline Bi (Rhom_A7) is surround by three neighbors with bond length ≈ 3.07 Å and another three neighbors with bond length ≈ 3.52 Å. Each bonded Bi-Bi atomic pair, i.e., with H-A indices 1001, have no common bonding atom. As a result, the Bi-centered coordination polyhedron (BIP) is indexed as $(6_0)$. Indices for BAPs and BIPs and their mole fractions (to the same species) in stable compounds ($Ba_2Bi$, $Ba_5Bi_3$, $Ba_4Bi_3$, $Ba_{11}Bi_{10}$, $BaBi_3$), crystalline Ba (BCC) and Bi (Rhom_A7) are listed in Table 1. The BIP in crystalline $Ba_2Bi$ is indexed as $(1_4,8_5)$ since each Bi is surround by one 1441 and eight 1551. The BIP in crystalline $Ba_4Bi_3$ is indexed as $(4_4,4_5)$ since each Bi is surround by four 1441 and four 1551. The BIP in crystalline $Ba_5Bi_3$ is indexed as $(3_4,6_5)$ since each Bi is surround by three 1441 and six 1551. The BAP in those compounds have high CNs, as listed in Table 1.

Aiming to gain more information, especially dynamic property, about the coordination polyhedrons, the lifetime of polyhedron has been calculated. The lifetime ($L_i$) for a polyhedron with a given type of indices is defined as the sum AIMD steps from its birth to its transformation to other types. Note that there are lots of transition entities with their lifetime of one or two steps; and these transition entities were ignored when we calculated the average lifetime $\langle L_i \rangle$.

## 3. Results and discussion
### 3.1 Energy analysis

The AIMD predicted enthalpy of mixing ($\Delta H_{mix}$) of molten $Ba_{1-x}Bi_x$ at 1123 K (Figure 1) agree well with the results by CALPHAD modeling using the associate model [10]. The enthalpy of formation ($\Delta H_{form}$) for the stable compounds predicted by DFT calculations at 300 K (using the GGA-PBE potential) [10] are also plotted in

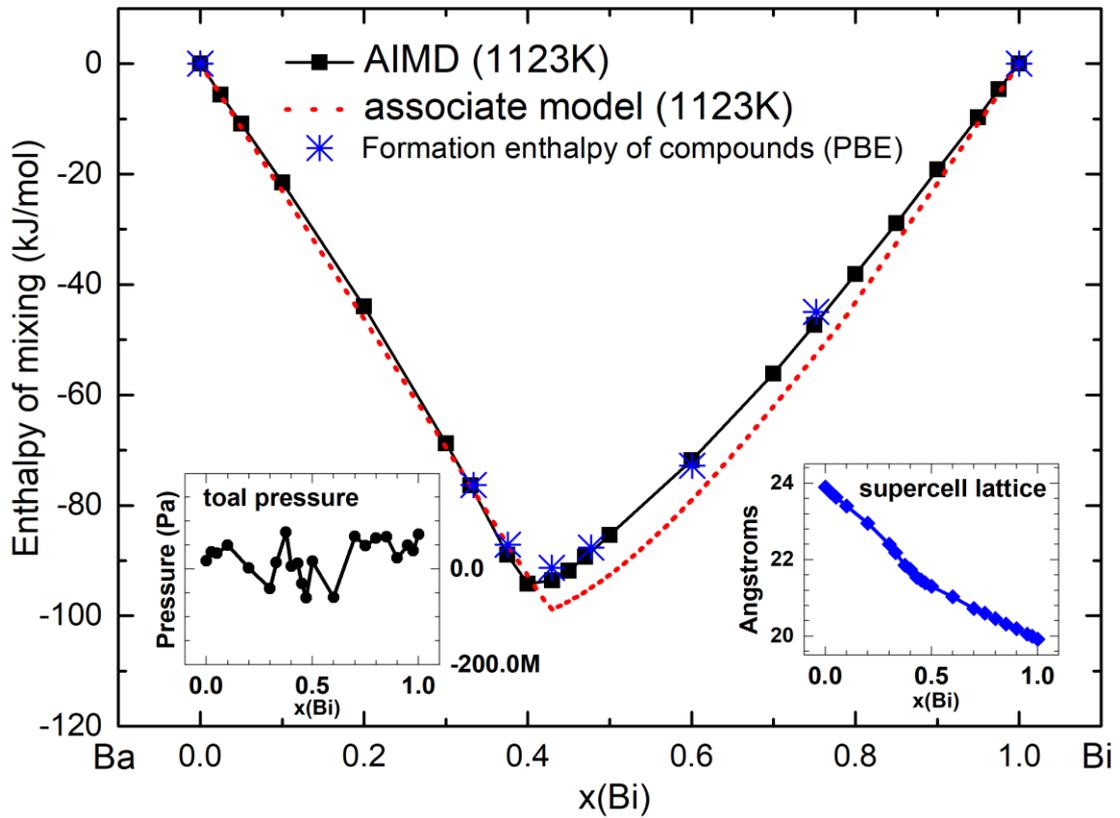

**Figure 1**. It is seen that $\Delta H_{form}$ is close to the AIMD predicted $\Delta H_{mix}$, and there is a deep valley on the $\Delta H_{mix}$ curve around the composition of compound $Ba_4Bi_3$, corresponding to the highest melting point in the Ba-Bi system, similarly to the Tl-Te [3], Pb-Te [6], and Ca-Sb [7] systems. In the following sections, various analysis methods are used to quantify the clusters in the liquid.

**3.2 Structure prediction: Ba-rich melt**

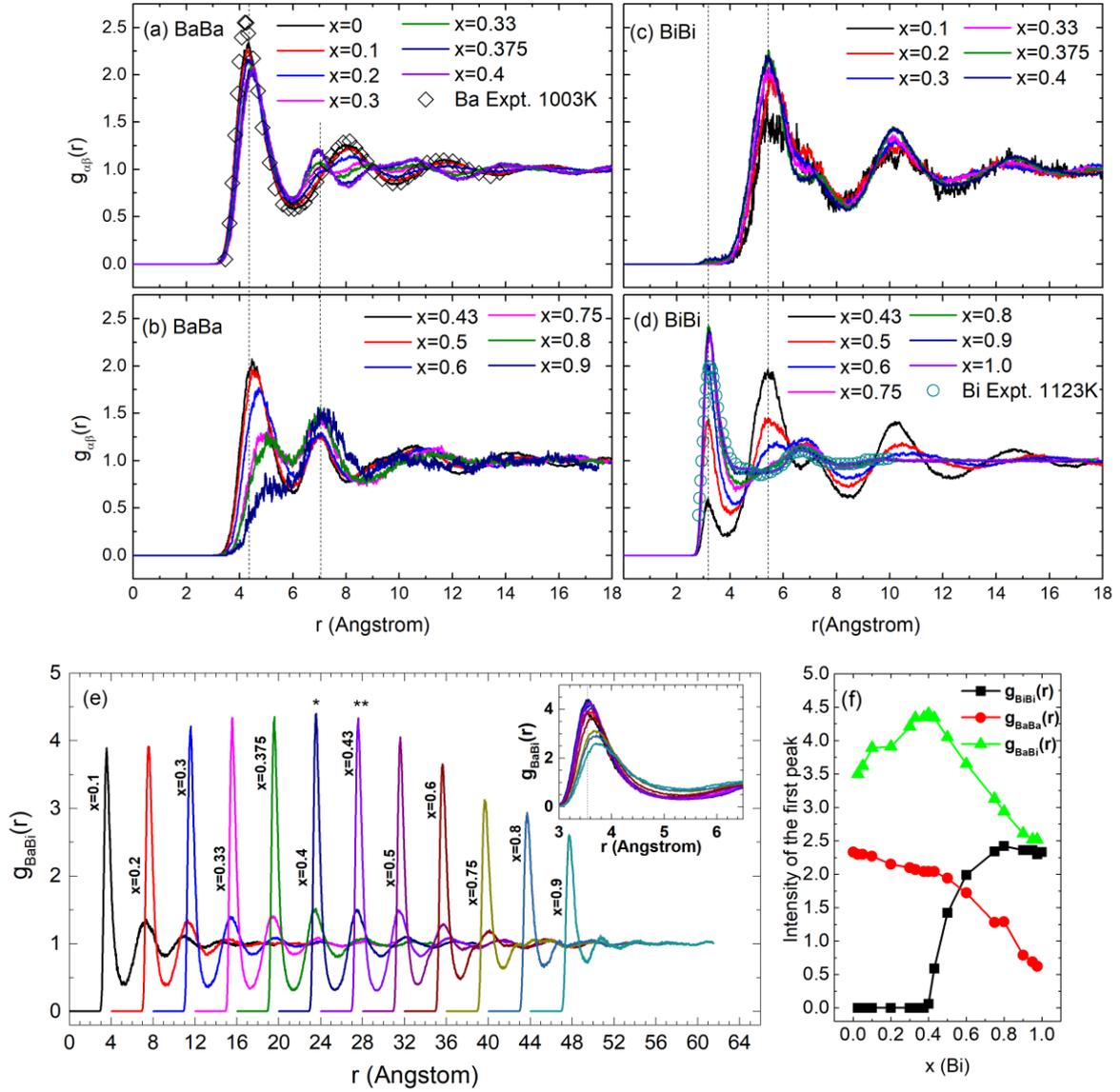

**Figure 2** shows the partial PCFs of molten Ba$_{1-x}$Bi$_x$ at 1123 K from the present AIMD simulations. According to

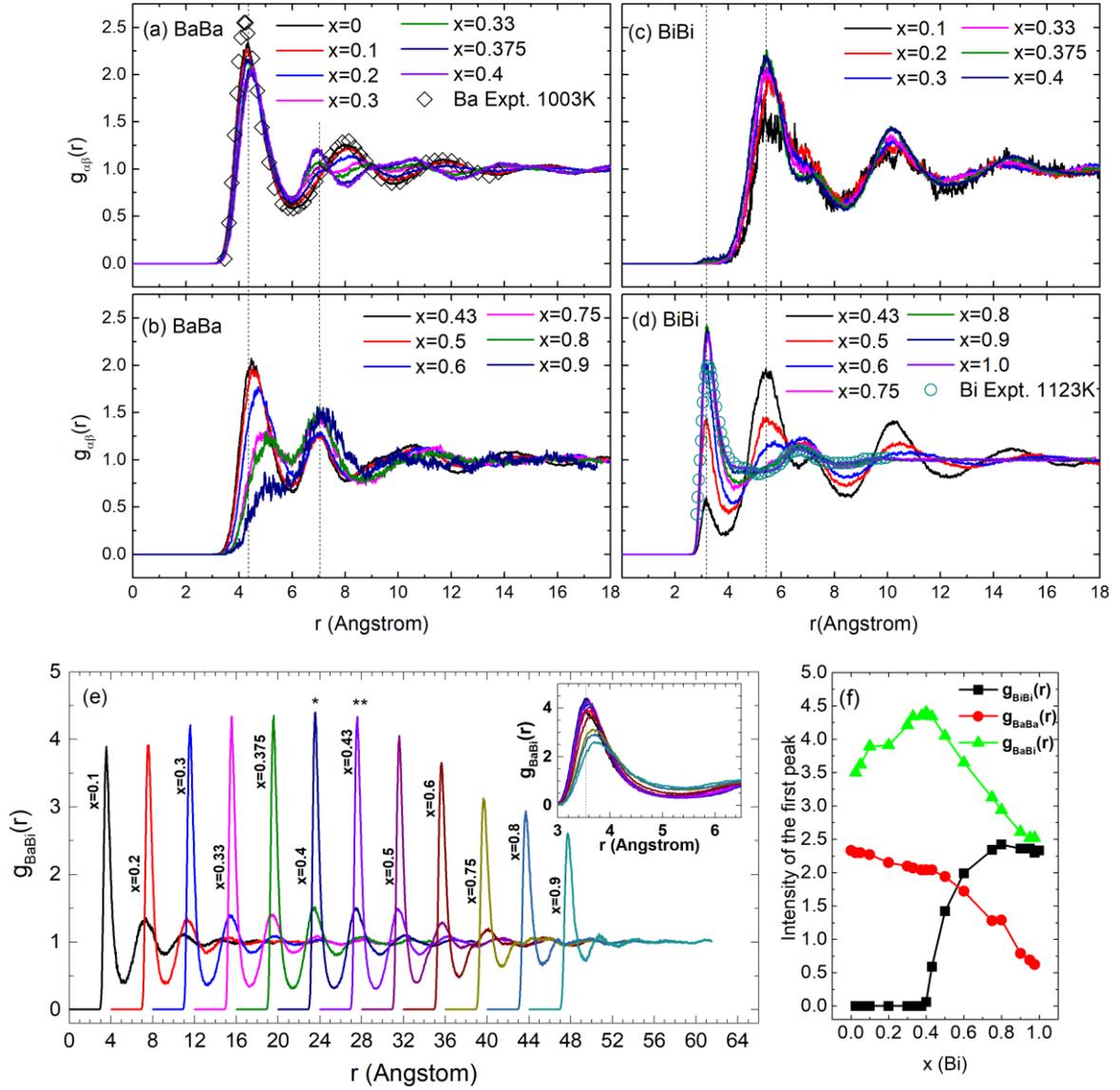

**Figure 2**f, the intensity of the first sharp peak of $g_{BaBi}(r)$ is stronger than those of the $g_{BaBa}(r)$ and $g_{BiBi}(r)$; indicating there is a strong chemical affinity between the unlike species. This fact is much pronounced when $x \leq 0.43$ (i.e., Ba-rich melt). The strong chemical affinity is also observed by the change of bond length, for example, the mean value (3.56 ± 0.02 Å) is shorter than the sum radii of Ba and Bi atoms (i.e., 4.30/2 + 3.25/2 = 3.775 Å).

The strong chemical affinity results in that the Bi atoms bond to Ba species exclusively when $x \leq 0.40$, evidenced by the absence of nearest neighbor peaks on $g_{BiBi}$-($r$) near 3.25 Å (

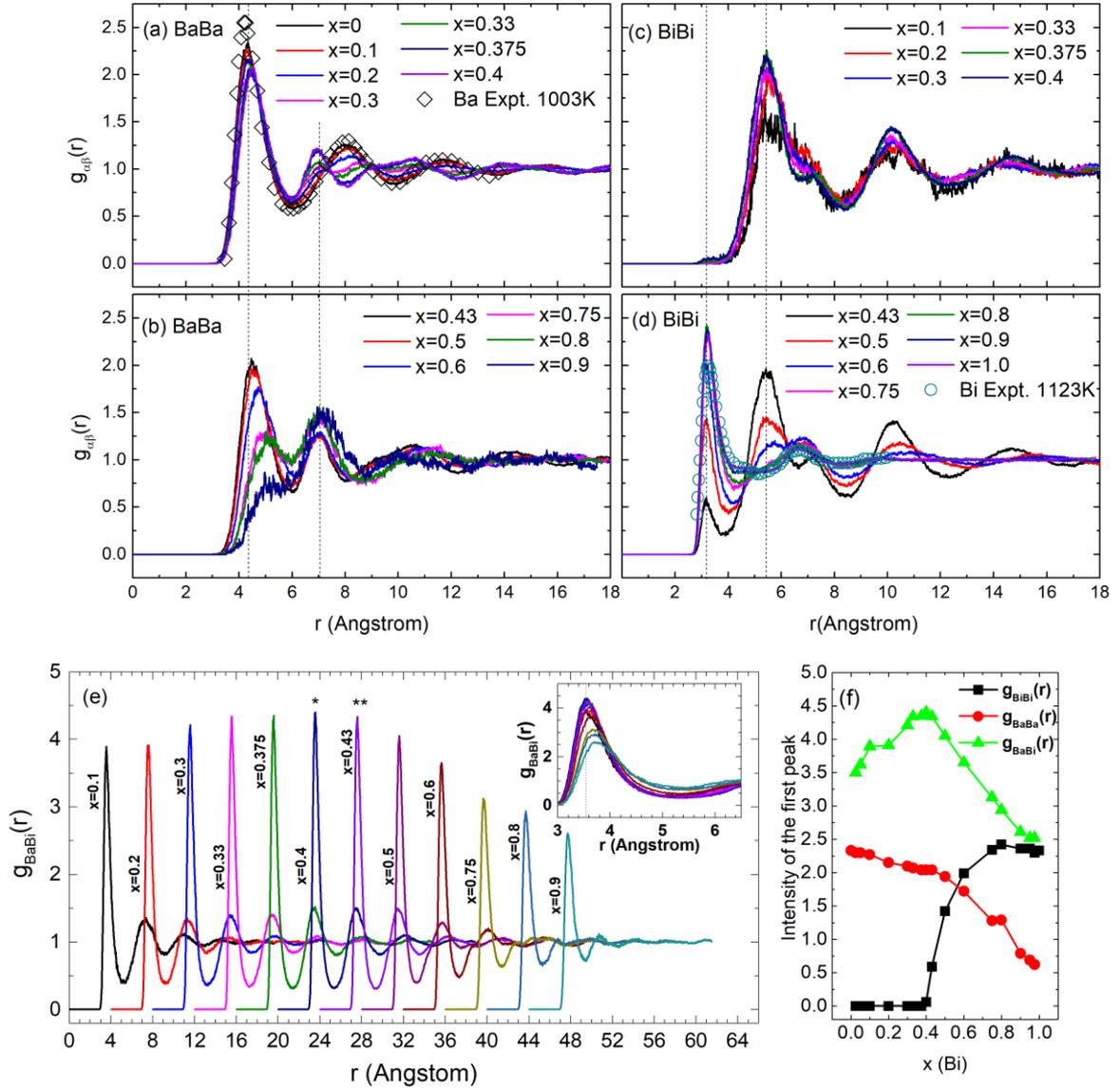

**Figure 2**c-d). This fact is also can be evidenced by the evolution of partial CNs, $Z_{BiBi}$, $Z_{BiBa}$, and $Z_{Bi}$, i.e., $Z_{BiBi} \approx 0$, $Z_{Bi}$ mainly contributed by $Z_{BiBa}$ when $x \leq 0.40$ (shown in Figure 3). A pronounced peak located at 5.45 Å on $g_{BiBi}(r)$, see

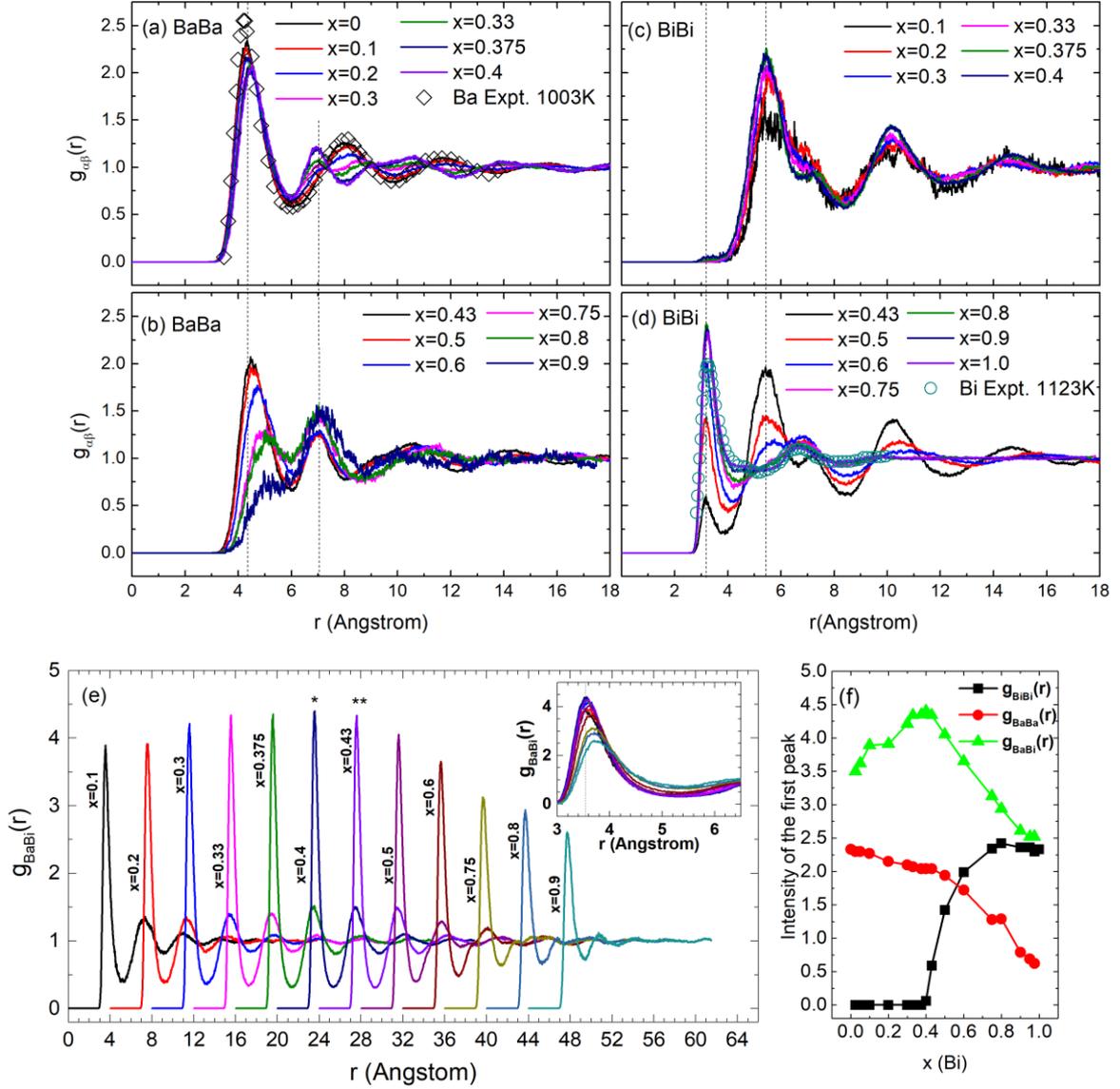

**Figure 2**c and d, is a distinct evidence regarding the connection of BIP.

Pre-peak on the left of the static SF indicates that there are large-scale repeated structures in the melt [43, 44], namely MRO. It has been reported that there is noticeable SF pre-peak for Ni rich melts (or amorphous) (e.g., Ni-Al [18, 45], Ni-Ti [23], Ni-Nb [23], Ni-Ta [23], and Ni-Pt [38]) and Al rich melts (e.g., Al-Ni [45, 46], Al-Co [45], and Al-Fe [47]). The pronounced peak located near q ≈ 1.37 Å$^{-1}$ on $S_{BiBi}(q)$ when $x$ = 0.10 and 0.43 is noticeable (Figure 4b and c). Their locations are on the left of the SF main peak of pure Bi (Figure 4d). These low $q$ peaks on $S_{BiBi}(q)$ are clear signals for the large-scale repeated structures for Bi in medium range, larger than the average bond length of Bi-Bi. The snapshots of the BIPs when $x$ = 0.025, 0.05, and 0.10, are shown in Figure 5a-c. They connect with each other through vertex, edge, face, and/or bipyramid sharing.

The peak located near 8.01 Å on $g_{BaBa}(r)$ (

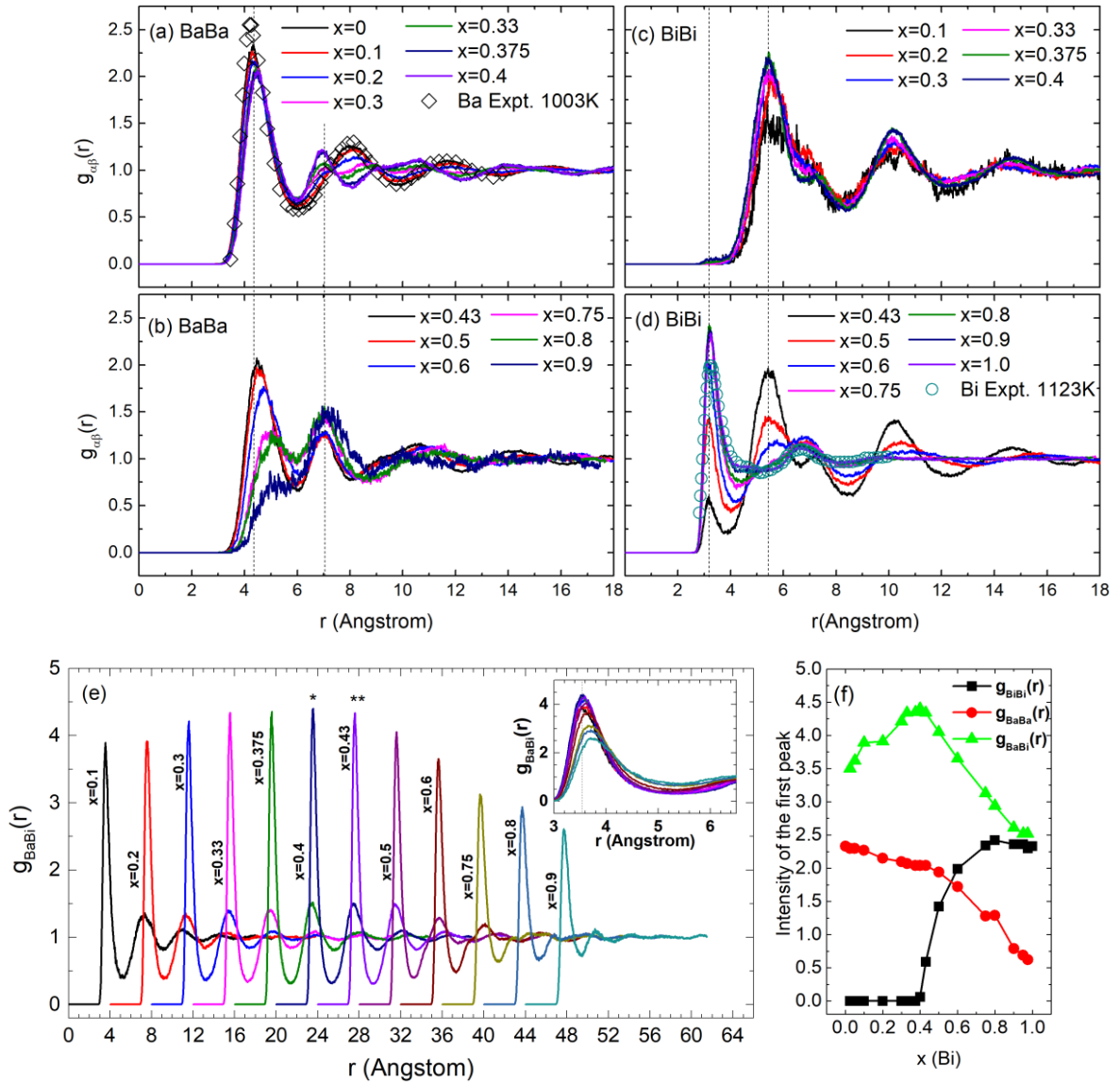

**Figure 2**a) can be seen as a signal for the free liquid Ba since there exists the same peak on the PCF of molten Ba. The Ba rich melt can be seen as the mixing of the connected BIPs and free liquid Ba. The weakening of this peak (

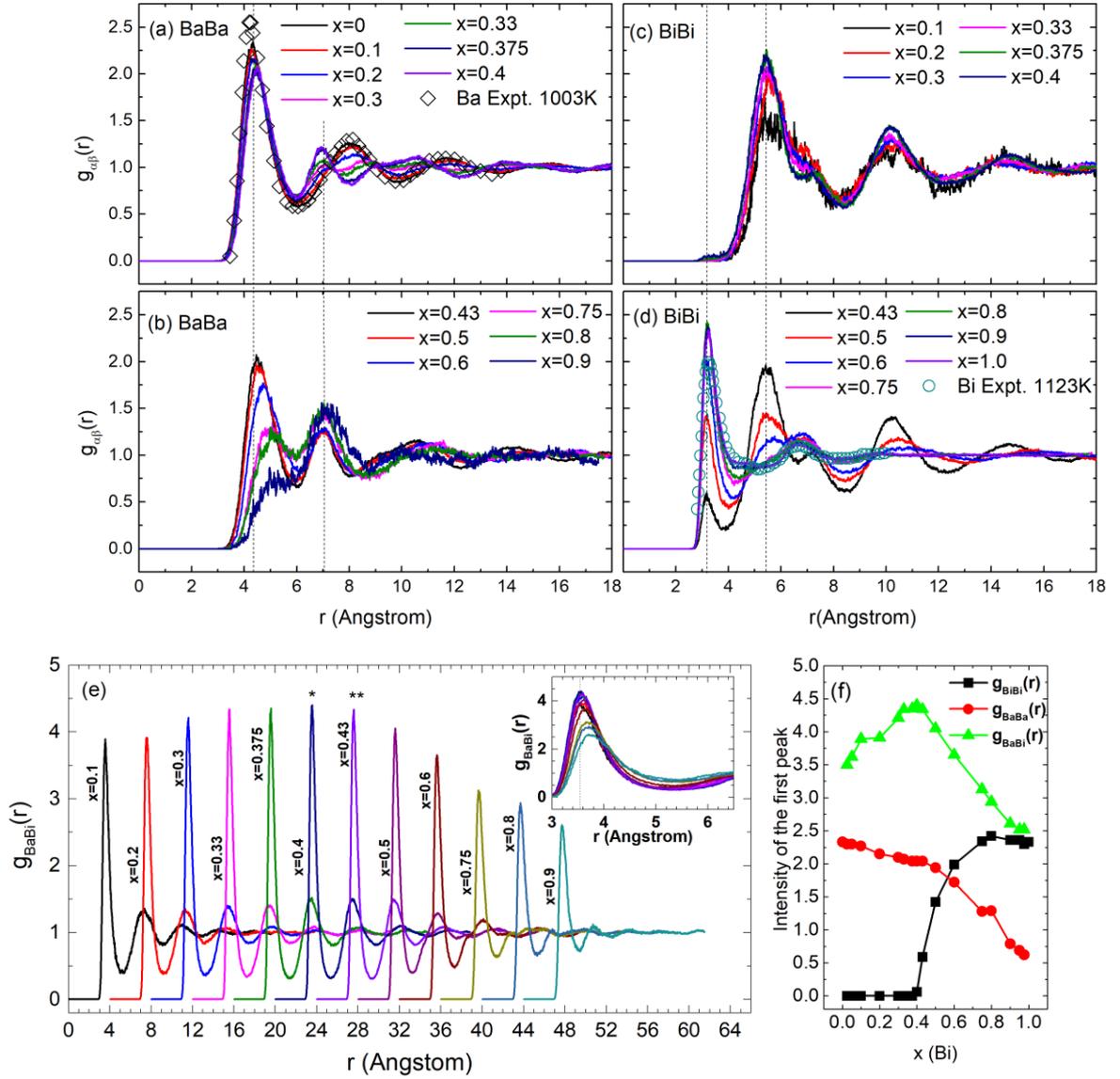

**Figure 2**a and b) results from the reduction of free liquid Ba. A peak near 1.4 Å$^{-1}$ $S_{BaBa}(q)$ is also noticeable for $x = 0.43$ (Figure 4c), indicating that this melt is highly ordered not only for Bi but also for Ba over a medium range.

### 3.3 Structure prediction: Bi-rich melt

The chemical affinity between the unlike species tends to decrease towards Bi-rich melt, based on the shift of the g$_{BaBi}$($r$) first sharp peak from 3.56 Å to 3.78 Å when $x >$ 0.43 and the decrease of its intensity with increasing $x$ (see the inset of

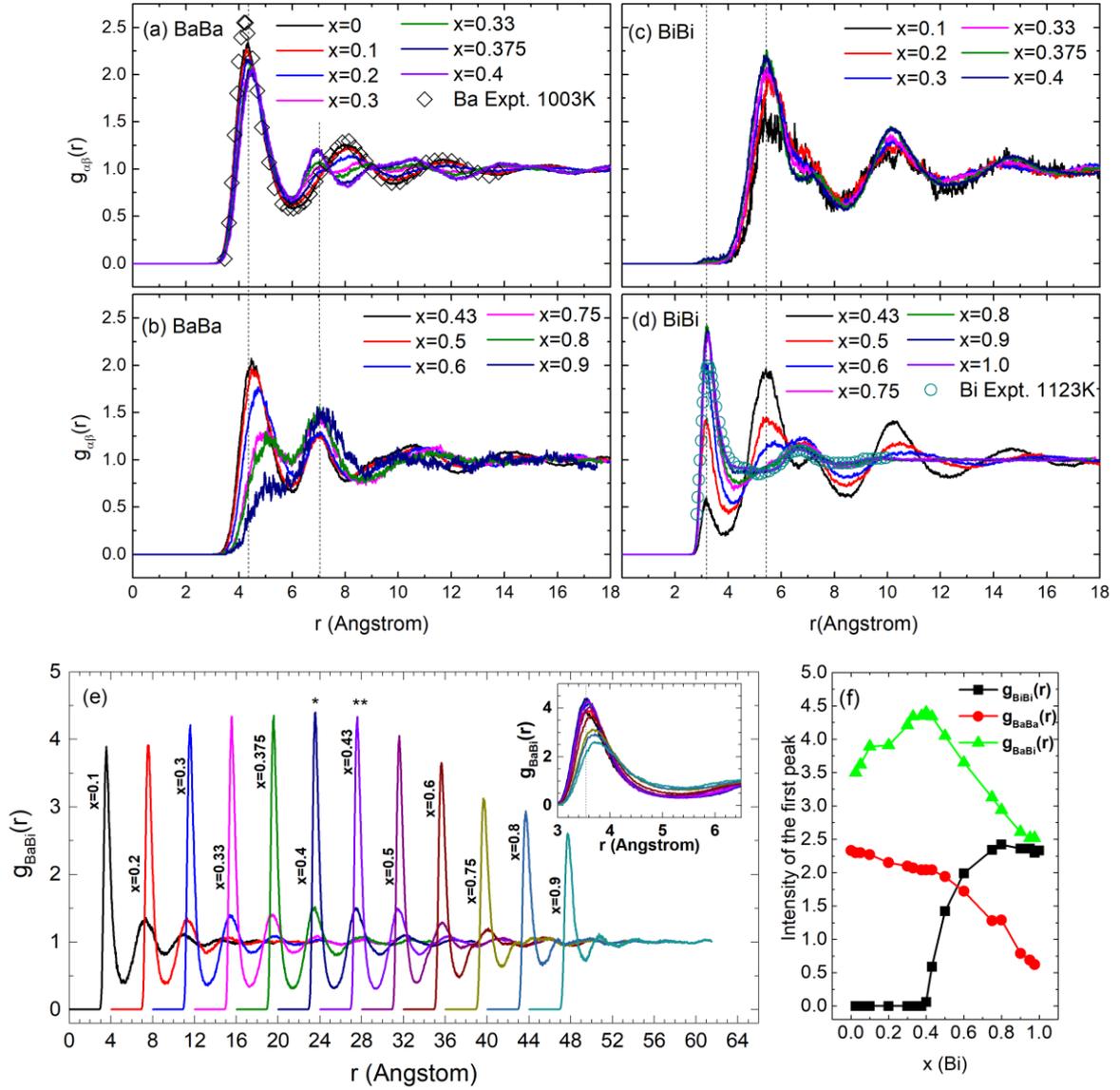

**Figure 2**e). In Bi rich melt, the chemical affinity between the unlike species is weaker than that in the Ba rich melt according to the weaker intensity of the $g_{BaBi}(r)$ first sharp peak, especially when $x \geq 0.7$ (see

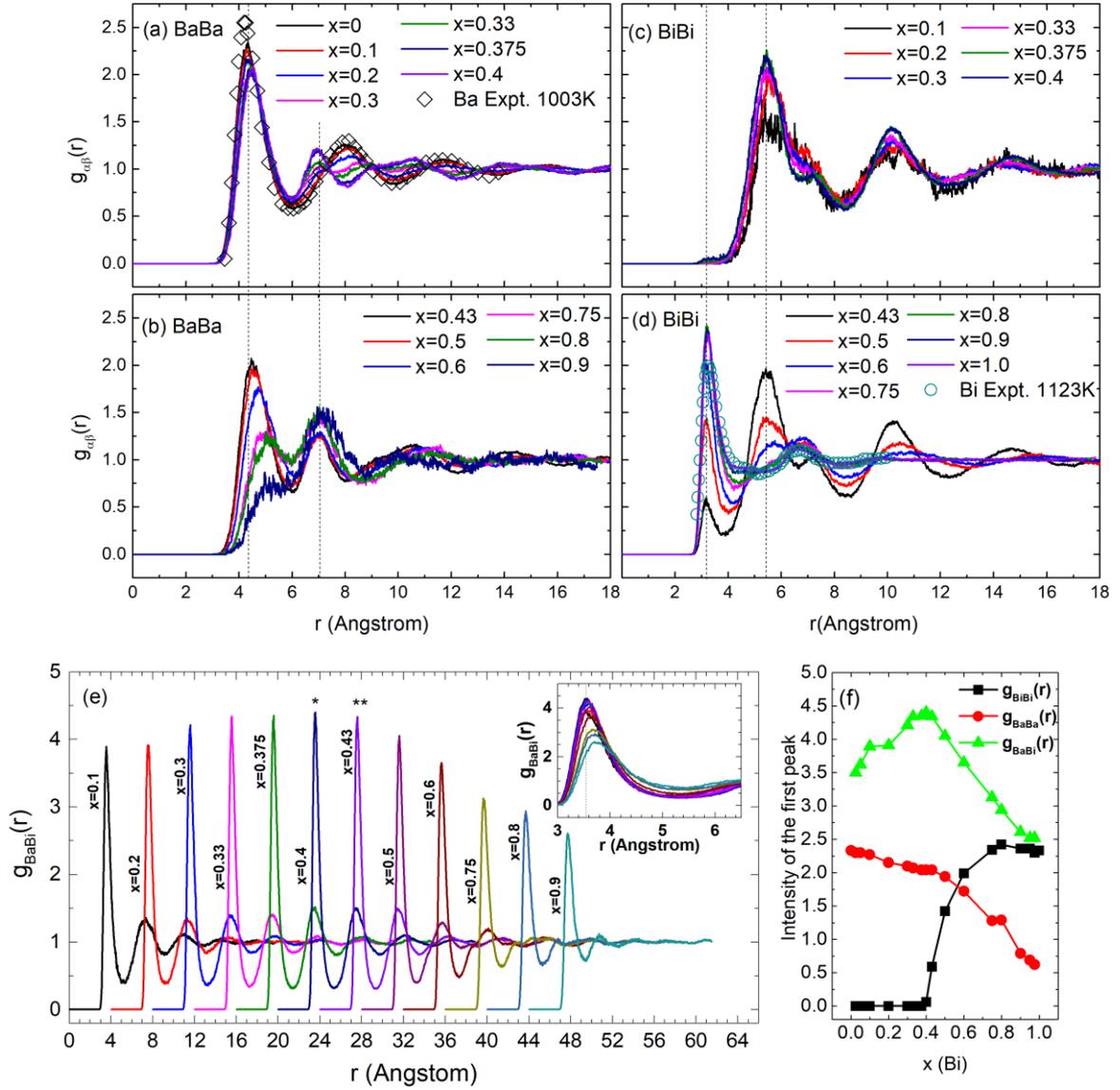

**Figure 2**e). However, the chemical affinity between the unlike species is still stronger than that between the like species. This fact results in Ba atoms surrounded by Bi. According to Figure 3, in the $Ba_{0.1}Bi_{0.9}$, $Ba_{0.05}Bi_{0.95}$, and $Ba_{0.025}Bi_{0.975}$ melts, $Z_{BaBa}$ is less than 1.0, and Ba atoms are mainly surrounded by Bi species with very large $Z_{BaBi}$ (close to or larger than 15.0). This feature is expected to be responsible for the extremely low activity of Ba in Bi rich melt, such as $8.7 \times 10^{-12}$ at $x = 0.95$ and $T = 1123$ K [1].

The pronounced PCF peak located at 7.01 Å of $g_{BaBa}(r)$ when $x \geq 0.75$, see

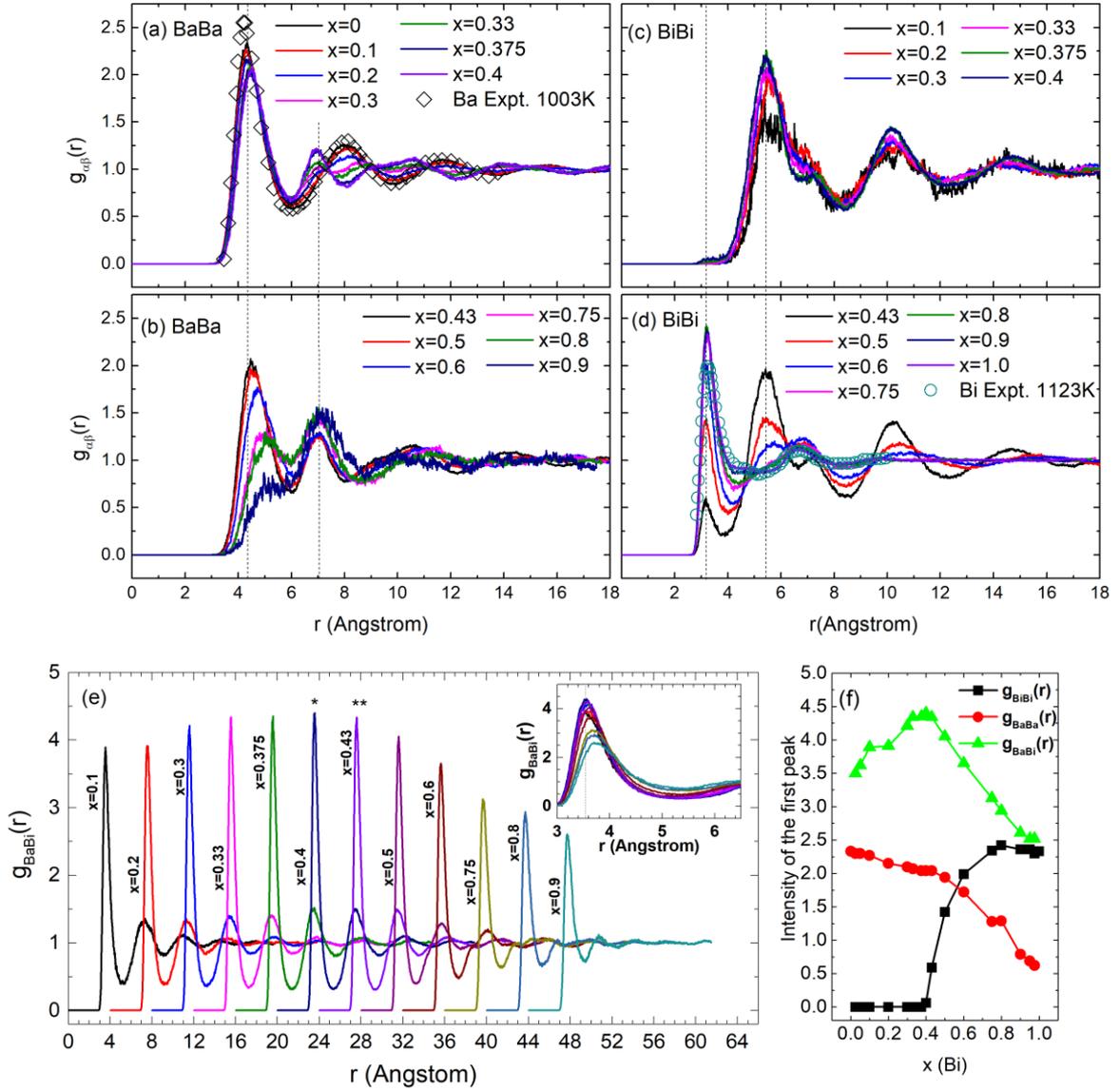

**Figure 2**b, is an evidence regarding the connection of BAP. The pronounced SF peaks located near 1.05 Å$^{-1}$ (*x* = 0.90) and 1.26 Å$^{-1}$ (*x* = 0.75) on $S_{BaBa}$(q) is also evidences of Ba atoms ordering in medium range (Figure 4e and f). The weak SF pre-peak on $S_{BiBi}$(q) (Figure 4f), near 1.05 Å$^{-1}$ (*x* = 0.75), indicates the ordering of Bi atoms in medium range to some extent. Figure 5d-f show the snapshots of BAPs when *x* = 0.90, 0.95, and 0.975. These BAPs are connected as predicted based on the analysis of PCFs. The Bi rich melt looks like the mixing of these BAP and the free liquid Bi.

## 3.4 Short range ordering and medium range ordering

*3.4.1 Bonded-pairs*

Figure 6a shows there is a maximum on the curve of the Ba-Ba rooted bonded-pair 1661 near $x = 0.43$. Conversely, there is a minimum near the same location on the bonded-pairs with the small H-A indices of 1421, 1422, 1321, 1311, 1301, and 1201 (see Figure 6e-g ). The combined fractions of the Ba-Ba bonded-pairs with large indices (1661, 1651, 1551, 1541, 1441, and 1431) have large values (> 0.70) in Ba rich melts, reaching a maximum at $x = 0.43$ and decreasing rapidly towards Bi rich melts at $x > 0.43$ (see Figure 6d). The sum of the Ba-Ba bonded-pairs with small indices displays a converse nature (see Figure 6h).

According to Figure 7a-c, the Ba-Bi rooted bonded-pairs of 1661, 1551, 1541, 1441, and 1431 are favored in Ba rich melts. The combined fraction of the bonded-pairs of 1661, 1651, 1551, 1541, 1441, and 1431, remains high (> 0.79) with a slightly decreasing trend when $x \leq 0.43$ (Figure 7d). When $x > 0.43$, the combined fraction decreases rapidly with increasing $x$ (Figure 7d). On the other hand, the combined fraction of the small indices of 1421, 1422, 1321, 1311, 1301, 1201, and 1101, increases rapidly (Figure 7h). The total number of the bonded-pairs with large H-A indices per supercell reaches a maximum value around 558 near $x = 0.43$ (Figure 7i), while the total number of the bonded-pairs with small indices pairs reaches a maximum value around 342 near $x = 0.75$ (Figure 7j). These two compositions are near the compositions of the two associates $Ba_4Bi_3$ and $BaBi_3$ assumed in the CALPHAD modeling [10].

Figure 8d shows that the combined fraction of 1421, 1422, 1321, and 1311 with Bi-Bi root pair reaches a maximum at composition close to BaBi$_3$. These bonded-pairs may be related to the decomposition of 1431 Bi-Bi bonded-pair, which is the only Bi-Bi rooted pair in the compound BaBi$_3$. Since the fractions of Ba-Ba and Ba-Bi rooted bonded-pairs in molten BaBi$_3$ do not match well with those in the crystalline state, it indicates the poor structure correlation between crystalline BaBi$_3$ and its molten state. According to Figure 8e-g, when Bi is molten, the bonded Bi-Bi atomic pairs have several common neighbors, represented by 1311, 1301, 1201, and 1101, accompanied by the increased coordination number and density [34]. The considerable amount of 1001 and 1101 bonded-pairs (Figure 8g) indicate that the structure of molten Bi is correlated to crystalline Bi to some extent. The fractions of the bonded-pairs with indices 1301, 1201, 1101, and 1001 decrease with decreasing $x$ (Figure 8f-j), resulting in fewer free Bi atoms in liquid.

*3.4.2 Coordination polyhedrons*

The top ten BIPs in molten Ba$_4$Bi$_3$ ($x$ = 0.43) are ($6_4,2_5$), ($4_4,4_5$), ($5_4,4_5$), ($2_3,4_4,2_5$), ($3_4,6_5$), ($1_3,4_4,3_5$), ($1_3,6_4,1_5$), ($1_3,5_4,1_5$), ($1_3,5_4,3_5$), and ($1_3,3_4,5_5$) in the order of decreasing fraction. It is believed that the BIP with index ($4_4,4_5$) is inherited from crystalline Ba$_4$Bi$_3$ since it has the same index as that in compound Ba$_4$Bi$_3$ (Table 1). The index ($3_4,6_5$) is the same as the BIP in compound Ba$_5$Bi$_3$ (Table 1), which is a tricapped trigonal prism polyhedron [16]. The polyhedron with index ($3_4,6_5$) is preferred when effective atomic size ratio between solute and solvent atoms, R$^*$, is close to 0.732 [16] (in Ba rich melt, R$^*$ ≈ 0.756). Similar structures have been reported in molten $Al_{107}Fe$ [48] and metallic glass $Ni_{81}B_{19}$ [16]. Furthermore, some aforementioned polyhedrons can transform from each other, e.g., $(5_4,4_5) \rightarrow (3_4,6_5) \rightarrow (1_3,5_4,3_5)$, $(1_3,5_4,3_5) \rightarrow (1_3,3_4,5_5) \rightarrow (6_4,2_5)$, and $(4_4,4_5) \rightarrow (6_4,2_5) \rightarrow (1_3,4_4,3_5)$. Hence, these BIPs are strongly associated to compounds $Ba_4Bi_3$ and $Ba_5Bi_3$.

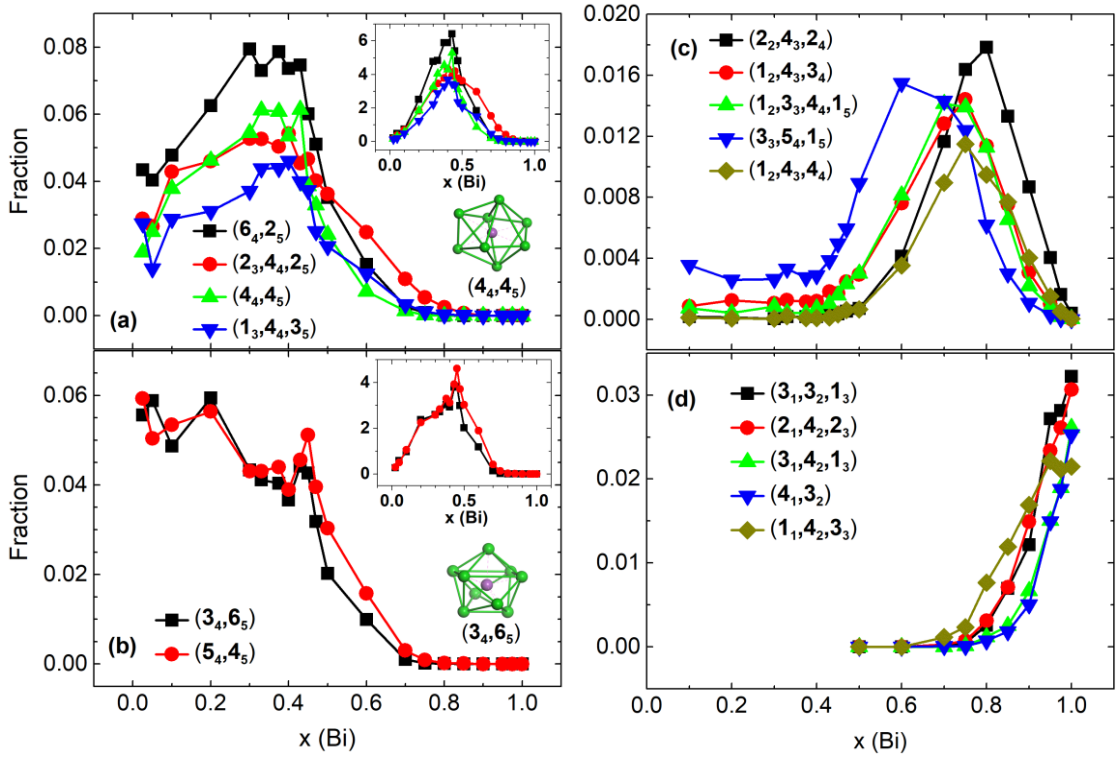

Figure 9a and b show the variation of the top six kinds of BIPs related to $Ba_4Bi_3$ with the indices of $(6_4,2_5)$, $(4_4,4_5)$, $(5_4,4_5)$, $(2_3,4_4,2_5)$, $(3_4,6_5)$, and $(1_3,4_4,3_5)$ as a function of Bi mole fraction ($x$). The number of these BIPs reaches a maximum at $x = 0.43$ (see the insets of

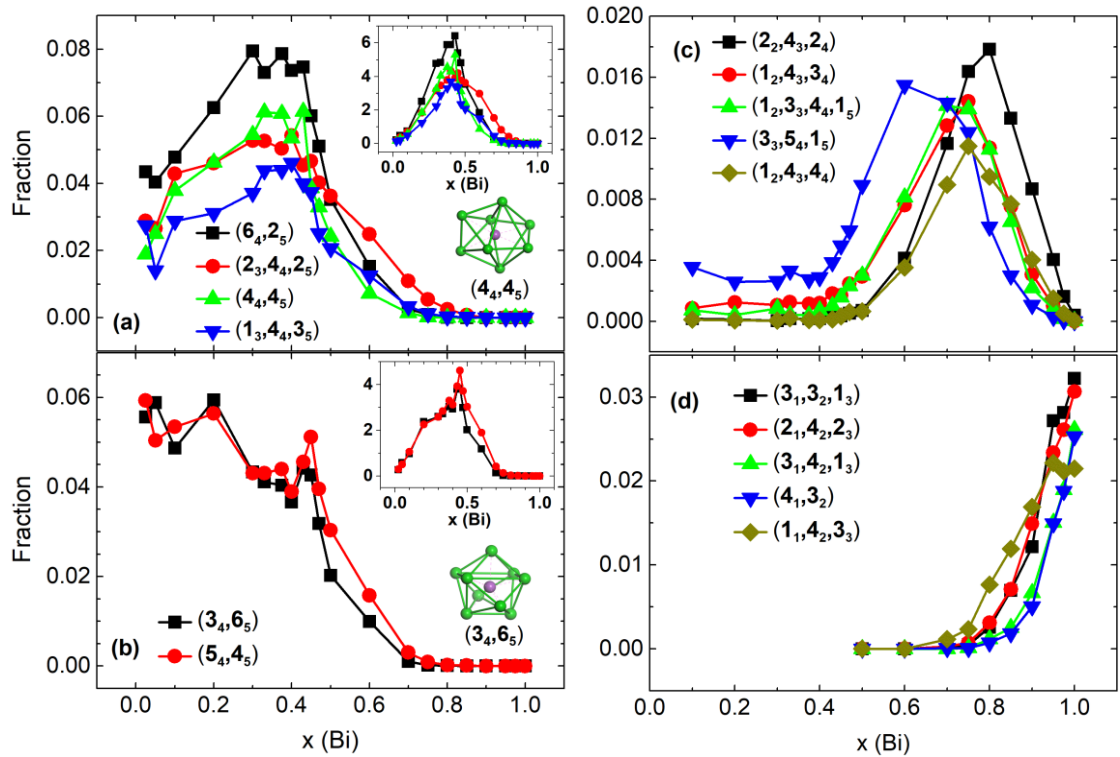

Figure 9a and b). With increasing *x*, the combined fraction of these six indices, shown in Figure 10, increases firstly, reaches a maximum at about 0.30 when $0.20 \leq x \leq 0.43$, and then decreases rapidly and disappear towards Bi rich melt ($x > 0.75$). The combined fractions of the aforementioned top ten BIPs related to molten $Ba_4Bi_3$ display a similar trend and reaches a maximum value at 0.45 when $0.20 \leq x \leq 0.43$. These BIPs associate with compound $Ba_4Bi_3$ and dominate in the Ba rich melt according to Figure 10.

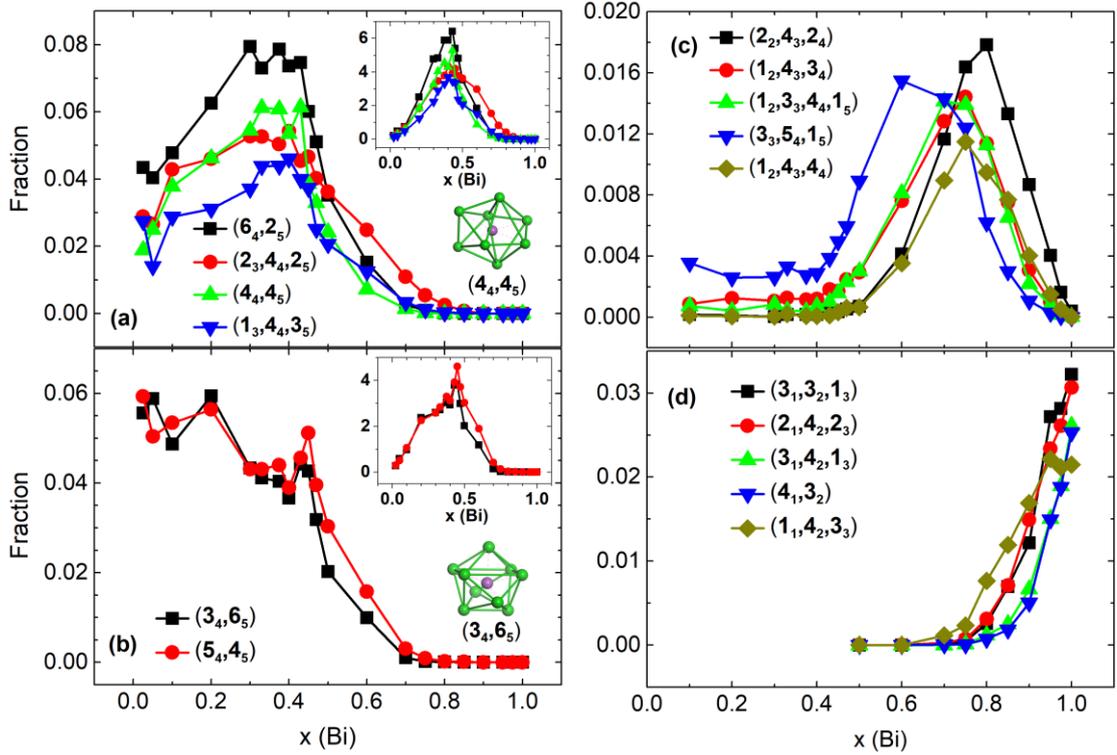

Figure 9c shows the variation of the top five BIPs related to molten BaBi$_3$ ($x = 0.75$) with the indices of ($2_2$,$4_3$,$2_4$), ($1_2$,$4_3$,$3_4$), ($1_2$,$3_3$,$4_4$,$1_5$), ($3_3$,$5_4$,$1_5$), and ($1_2$,$4_3$,$4_4$) as a function of $x$. Clearly, the fraction of each index reaches a maximum near $x = 0.75$ and their combined fraction reaches maximum at $x = 0.75$ (Figure 10). This fact is also found for the combined fraction of those BIPs with the top nine indices of molten BaBi$_3$, i.e., ($2_2$,$4_3$,$2_4$), ($1_2$,$4_3$,$3_4$), ($1_2$,$3_3$,$4_4$,$1_5$), ($3_3$,$5_4$,$1_5$), ($1_2$,$4_3$,$4_4$), ($2_2$,$4_3$,$3_4$), ($1_2$,$3_3$,$3_4$,$1_5$), ($3_2$,$4_3$,$1_4$), and ($1_2$,$5_3$,$2_4$,$1_5$). It is expected that these polyhedrons are special for molten BaBi$_3$. However these popular BIPs in molten BaBi$_3$ are greatly different from ($8_4$,$4_6$), the BIP in crystalline BaBi$_3$ (Table 1.). Moreover, Bi-centered ($8_4$,$4_6$) is not found in molten BaBi$_3$. Hence, the structure of molten BaBi$_3$ is weakly correlated to its crystalline state.

The top five popular polyhedrons in molten Bi are ($3_1$,$3_2$,$1_3$), ($2_1$,$4_2$,$2_3$), ($3_1$,$4_2$,$1_3$), ($4_1$,$3_2$), and ($1_1$,$4_2$,$3_3$). Each of them contains at least one 1101 pair. Some polyhedrons, such as $(1_0,2_1,4_2)$, $(1_0,2_1,3_2)$, and $(2_0,2_1,2_2)$, even contain the 1001 pair. These facts indicate that the structure of molten Bi at 1123 K is still correlated to its crystalline state (Rhom_A7) to some extent. According to Figure 9d, the amount of Bi-centered $(3_1,3_2,1_3)$, $(2_1,4_2,2_3)$, $(3_1,4_2,1_3)$, $(4_1,3_2)$, and $(1_1,4_2,3_3)$ increases with increasing Bi concentration, implying the presence of free Bi like liquid in Bi rich melt.

The BAPs are much more diverse than the Bi-centered ones, resulting in a very low fraction (less than 0.01) of each BAP. For example, the top BAPs in molten Ba, $Ba_2Bi$, $Ba_5Bi_3$, $Ba_4Bi_3$, $BaBi_3$, and $Ba_{0.05}Bi_{0.95}$ are $(1_3,5_4,5_5,2_6)$, $(1_3,6_4,5_5,2_6)$, $(1_3,6_4,5_5,3_6)$, $(4_4,8_5,4_6)$, $(1_2,4_3,6_4,4_5,1_6)$, and $(3_2,8_3,4_4)$ with fractions of 0.0089, 0.0064, 0.0052, 0.0050, 0.0029, and 0.0047, respectively. The combined fraction of the top ten BAPs is much lower than that of the BIPs (Figure 10). This fact is much more pronounced in the Ba rich melt. The maximum value of the latter is more than 0.45 when $x = 0.43$, while the maximum value of the former is less than 0.09 in Ba. The top five ones in Ba are $(1_3,5_4,5_5,2_6)$, $(5_4,6_5,2_6)$, $(1_3,6_4,5_5,1_6)$, $(1_3,4_4,7_5,1_6)$, and $(1_3,5_4,5_5,1_6)$ with fractions of 0.0089, 0.0085, 0.0082, 0.0080, and 0.0080, respectively. The top five BAPs in $Ba_{0.05}Bi_{0.95}$ are $(3_2,8_3,4_4)$, $(3_2,8_3,5_4)$, $(2_2,7_3,5_4,1_5)$, $(4_2,8_3,4_4)$, and $(2_2,10_3,4_4)$ with fractions of 0.0047, 0.0045, 0.0042, 0.0040, and 0.0040, respectively. All of them are much lower than the popular BIPs in the Ba rich melt, such as $(6_4,2_5)$, $(4_4,4_5)$, $(5_4,4_5)$, and $(3_4,6_5)$; see Figure 9a and b. Those BAPs inherit some features of the compounds (Table 1.), such as the high coordination numbers (16-18 for compounds, about 13-16 for melts, Figure 3) and the complex indices, but the structures of them are poorly associated with each other.

It is seen that the mean lifetimes of BIPs, no matter for all or the top popular ones, are about two times longer than those of the BAPs (see Figure 11). In addition, the BIPs in Ba rich melt have longer lifetimes than those in the Bi rich melt. The top popular polyhedrons have longer lifetimes than those of the other polyhedrons for both the BIPs and BAPs. These results indicate that (i) the BIPs in Ba rich melt are more stable than those in Bi rich melt; and (ii) the BIPs are more stable than the BAPs no matter in Ba rich or in Bi rich melt.

Figure 11 also shows that the BIPs and BAPs possess long lifetimes when composition is close to $Ba_4Bi_3$. It becomes much pronounced for the most popular BIPs. The mean lifetime for the top BIPs in molten $Ba_4Bi_3$ is about 31.8 steps, i.e., 95.4 fs. The BIPs of $(4_4,4_5)$ and $(3_4,6_5)$, associated with compounds $Ba_4Bi_3$ and $Ba_5Bi_3$ as above discussed, have the mean lifetimes about 37.2 and 39.6 steps (111.6 and 118.8 fs), respectively. They are the polyhedrons with the longest lifetimes.

*3.4.3 Medium range ordering*

As mentioned in Sec. 3.2, the pronounced peak on $g_{BiBi}(r)$ with a maximum located at 5.45 Å (see

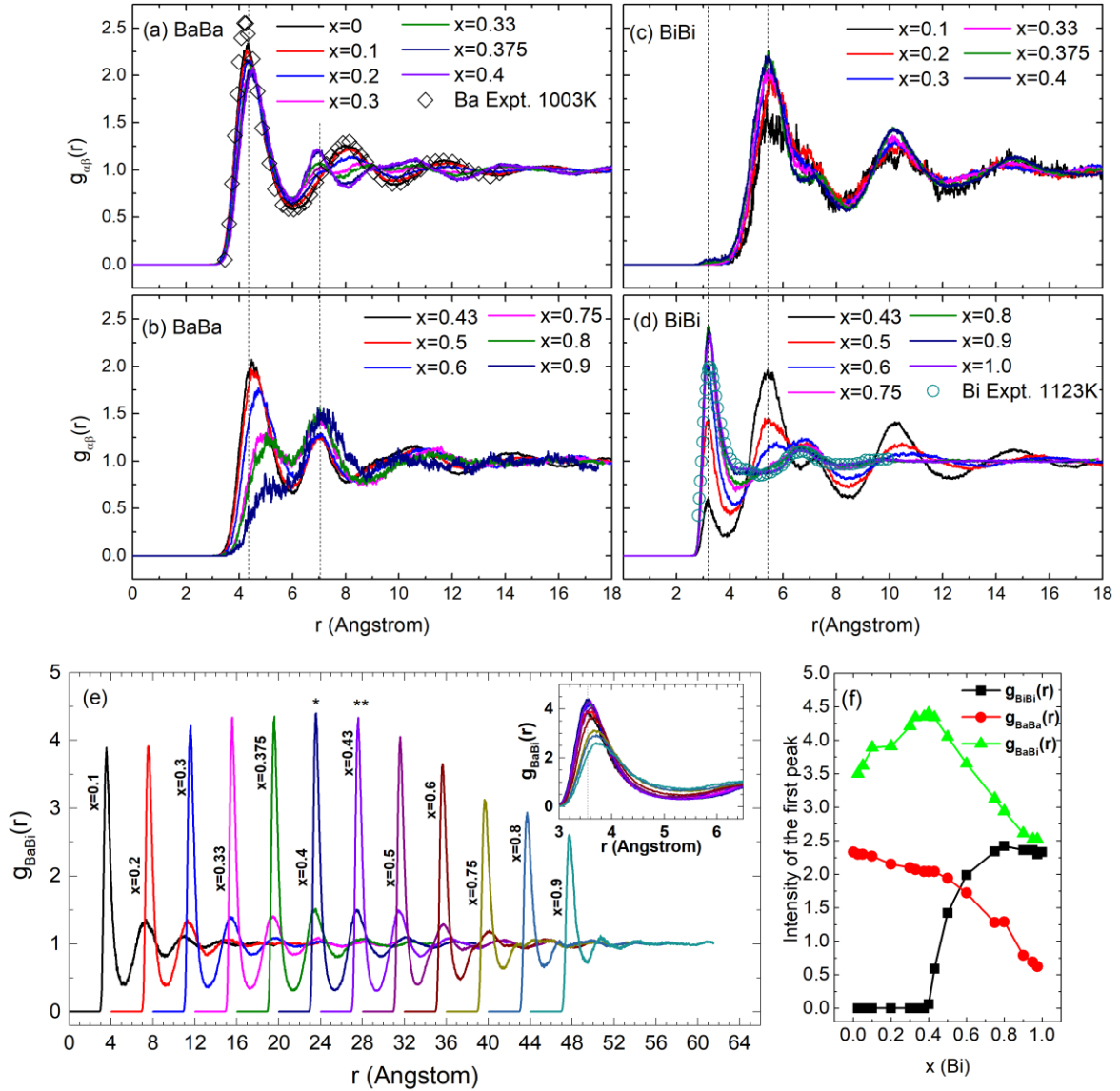

**Figure 2**a) when $x \leq 0.40$, is a signal for the connection of BIPs with vertex, edge, or face sharing. This connection was also reported in glass $Al_{75}Ni_{25}$ [16] and liquid/glass Cu-Zr [17]. Part of BIPs are connected with bipyramids sharing, such as pentagonal, tetragonal, or trigonal bipyramids, evidenced by the high fraction (reaching 0.60 at $x = 0.40$) of Bi-Bi rooted 1441 in Ba rich melt (Figure 8b).

According to

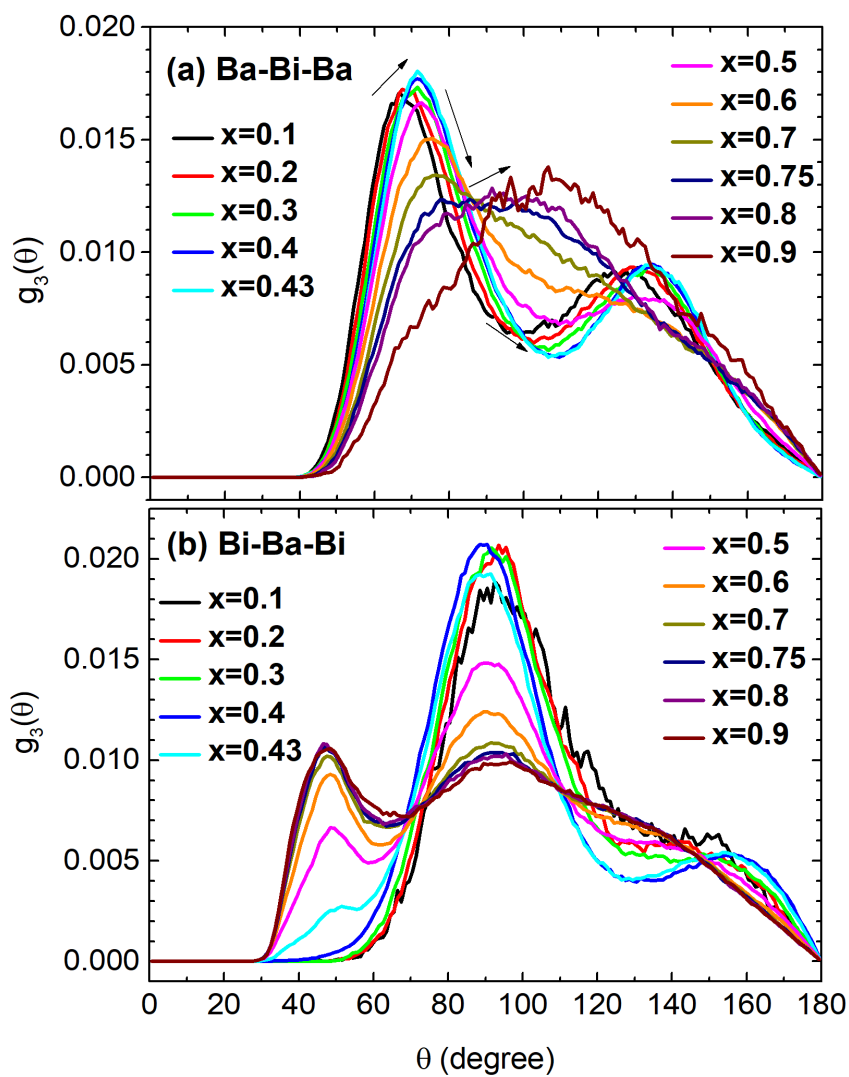

**Figure 12**a, in molten Ba$_{0.9}$Bi$_{0.1}$, the Ba-Bi-Ba triple prefers forming two kinds of bond-angles, 67.4° and 127.5°. Both of them slightly shift to large angle with increasing *x*. It is no doubt that the Ba-Bi-Ba triples are building blocks of the BIPs. In Ba rich melt, the shared Ba atom by BIPs and its neighboring Bi atoms (Bi-Ba-Bi triple) prefer forming right angle (90°) with Ba at the vertex (see

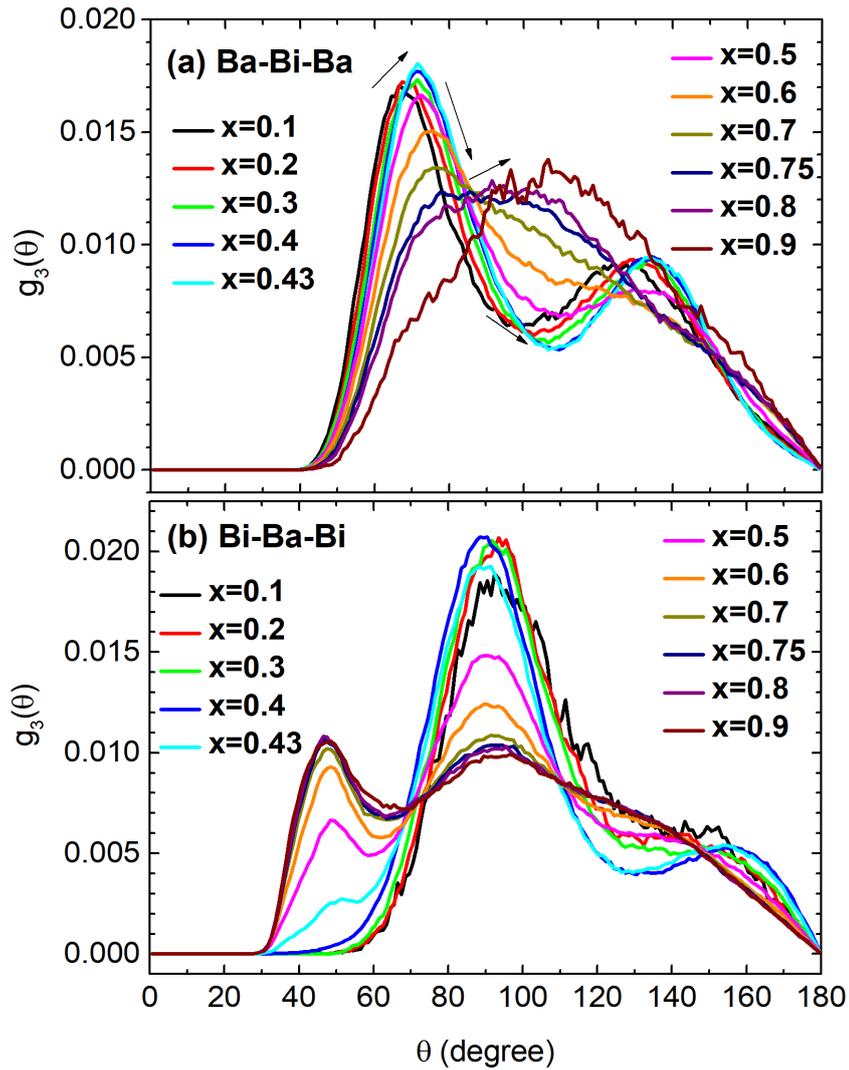

**Figure 12**b). As a result, the Ba-Bi-Ba-Bi-Ba combinations in the connected BIPs are expected to exist over medium range beyond short range.

As introduced in Sec 3.3, in the Bi rich melt, the BAPs are also connected with each other, evidenced by the pronounced peak on the PCF curve of $g_{BaBa}(r)$ located near $r = 7.13$ Å (

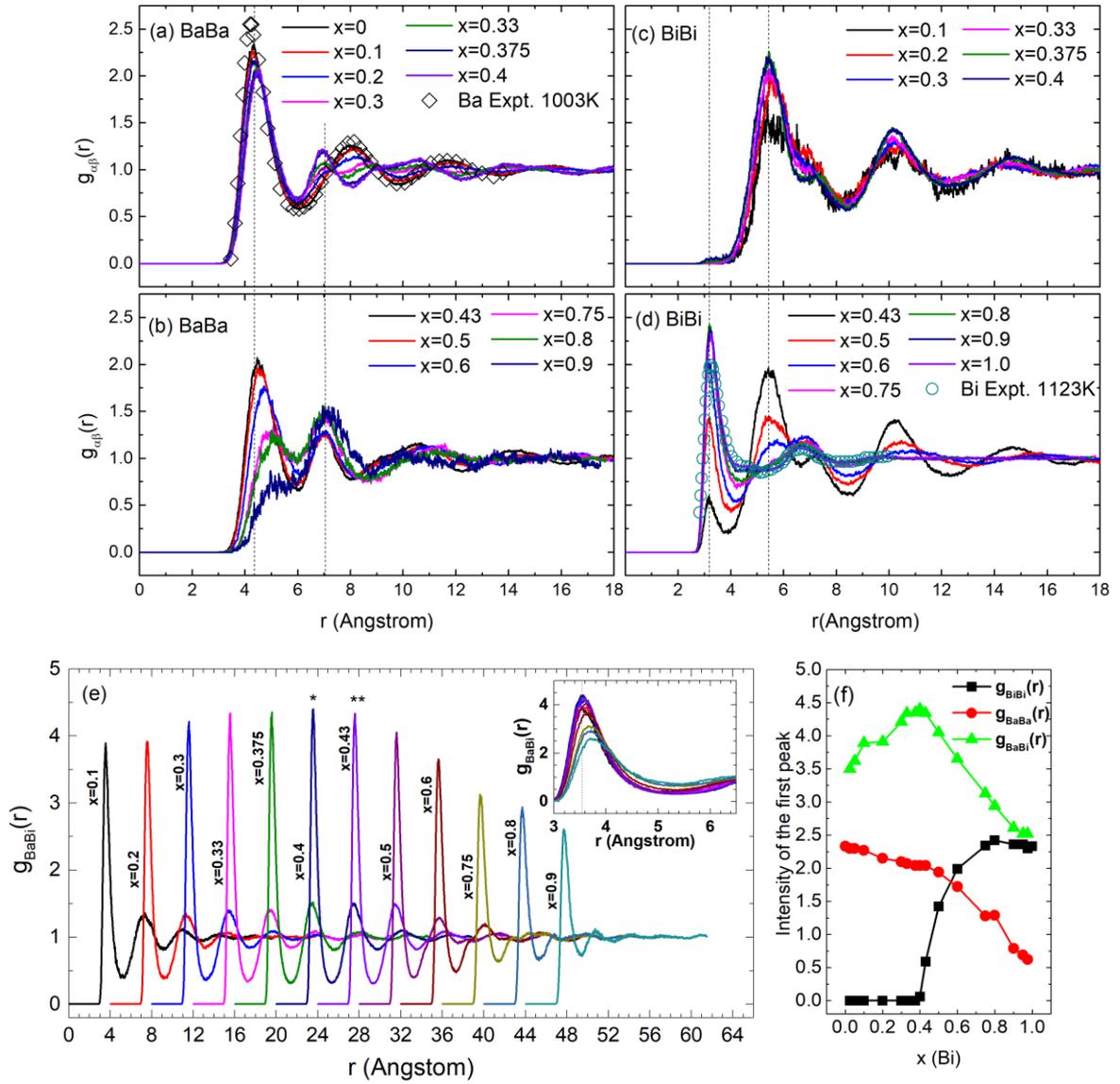

**Figure 2**b) and the snapshots of the BAPs (Figure 5d-f). According to

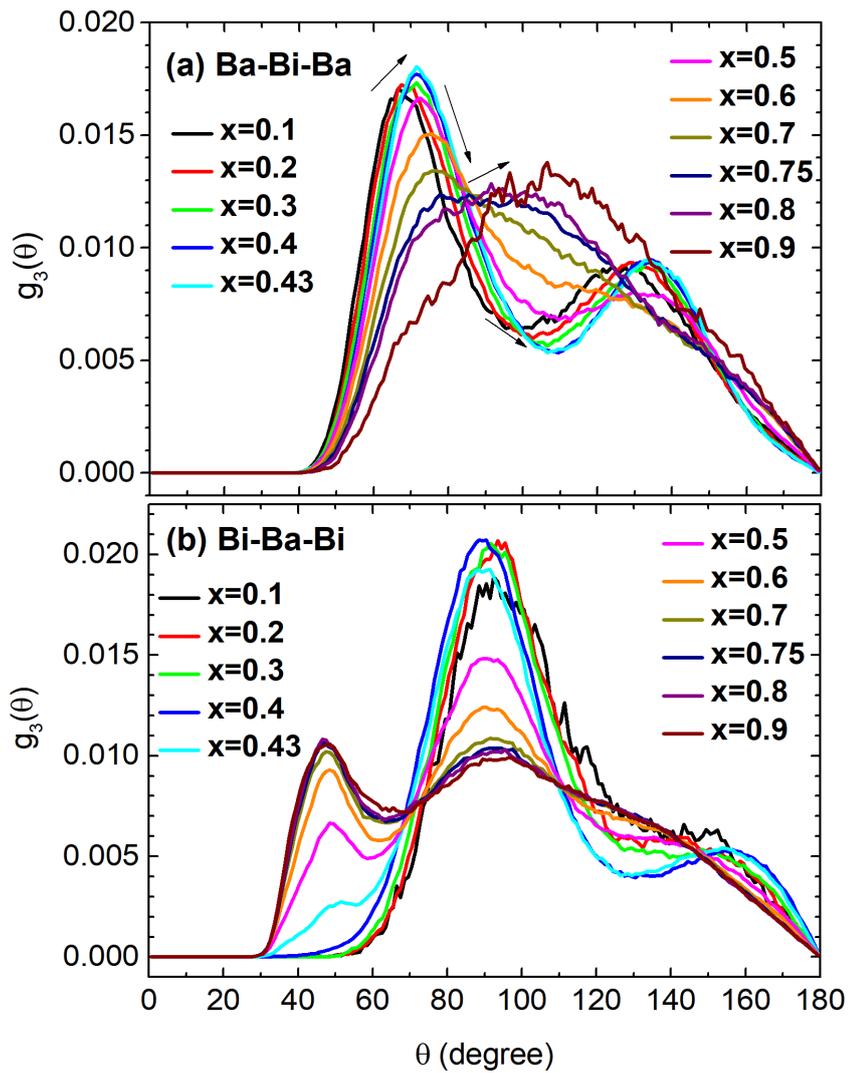

**Figure 12**b, in the Bi rich melt, the Bi-Ba-Bi triple prefer forming bond angles of 47.6° and 93.6°. The Ba-Bi-Ba triple prefer forming an angle of 105.7° (

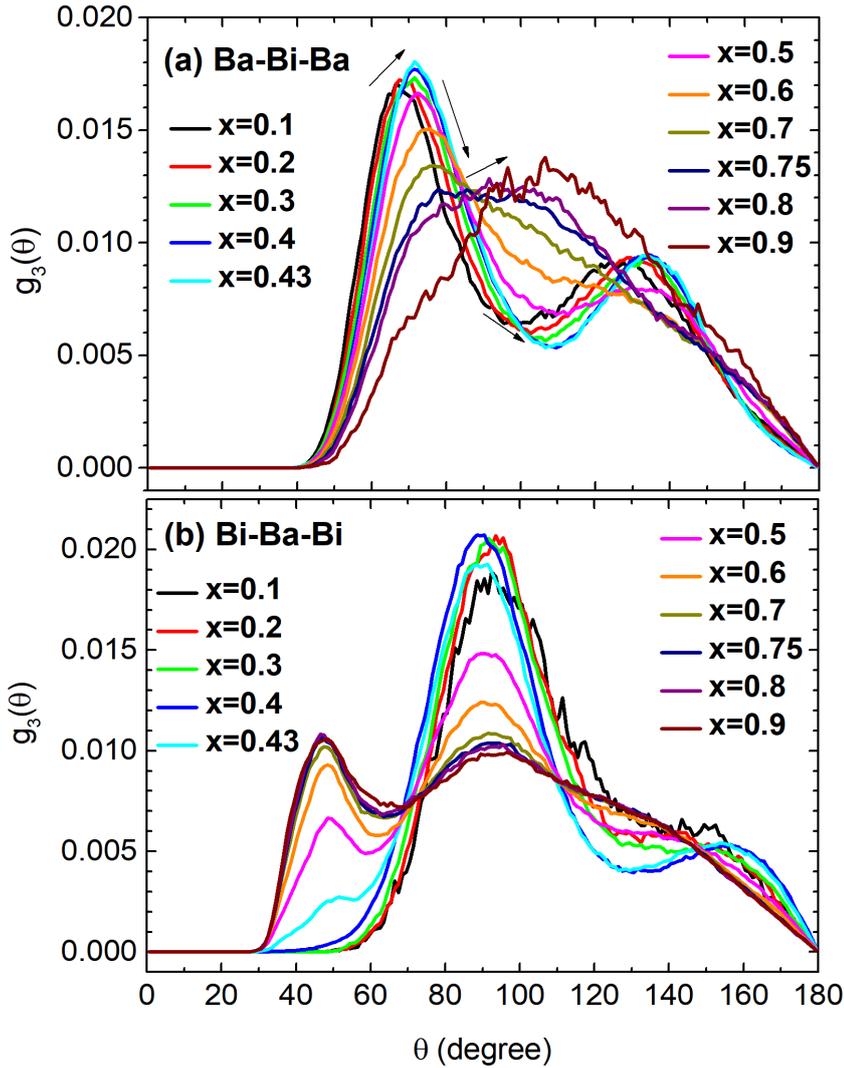

**Figure 12**a). So, it is expected that the Bi-Ba-Bi-Ba-Bi combinations in the connected BAPs are ordering beyond short range, indicating medium range ordering.

### 3.5 Associates

The present AIMD results, as above stated, indicate that the Ba rich melt can be simplified as a mixture of liquid Ba and the connected BIPs, the Bi rich melt can be simplified as a mixture of liquid Bi and the connected BAPs. The findings of medium range ordering in the connected BIPs and the connected BAPs support the hypothesis of fictive associates used to model the Ba-Bi system [10].

In the associate model for thermodynamic modeling, it is assumed that the associate has an exact stoichiometry [2, 4]. However, our computational results find that the mean Ba/Bi atomic ratios in the connected BIPs are roughly 8.05:1, 7.86:1, 6.24:1, 3.73:1, 7:3, 2:1, 5:3, and 3:2, when $x$ = 0.025, 0.05, 0.10, 0.20, 0.30, 0.33, 0.375, and 0.40, respectively. In addition, we also find that, as aforementioned, (i) the combined number of high indices of Ba-Bi bonded-pairs reaches the maximum near $x$ = 0.43 (Figure 7i), which are the main building blocks of the connected BIPs and dominate in the Ba rich melt; (ii) the numbers of the specific BIPs related to compound $Ba_4Bi_3$ reach the maximum near $x$ = 0.43 (Figure 9); and (iii) the BIPs are related to compound $Ba_4Bi_3$ dominating in Ba rich melts (Figure 10). Though the assignment of exact stoichiometry seems an arbitrary choice in associate model (e.g., $Ba_4Bi_3$ associate in Ba rich melt), according to facts evidenced from the current AIMD simulation, just stated, it is a matter of cause, or the best choice, to name the associate with a fixed composition of $Ba_4Bi_3$.

The mean Bi/Ba atomic ratios in the connected BAPs are roughly 4:1, 5.55:1, 8.15:1, 11.77:1, and 13.66:1 when $x$ = 0.80, 0.85, 0.90, 0.95, and 0.975, respectively. Obviously, they are also not presented as stoichiometric compounds in the Ba-Bi system. However, the AIMD results indicate that (i) the combined number of low indices of Ba-Bi bonded-pairs, the main building blocks of the connected BAPs, reaches the maximum near $x$ = 0.75 (Figure 7j); and (ii) the combined fraction of the BIPs, especially for $BaBi_3$ melt, being shared by BAPs in the connected BAPs, reaches a maximum near $x$ = 0.75 (Figure 10). Therefore, it is reasonable to assign additional

BaBi$_3$ associate to describe Bi-rich melt, as proposed by Liu et al. [10]. It is noted that this associate is weak ordering and poorly correlated to compound BaBi$_3$.

These structural features found in the Ba-Bi liquid are expected to occur in other liquids with strong chemical affinity between unlike species. For example of an A-B binary with A rich MRO in A rich melt and B rich MRO in B rich melt, these MROs can be the connections of the solute-centered polyhedrons [16] or the solute-solvent network [16, 18]. In the Al-Ni system, it was reported that there are string-like MRO in Al rich side (in glass Al$_{75}$Ni$_{25}$) [16] and solute-solvent network MRO in Ni rich side (in molten Ni$_3$Al) [18]. For the Al-Co system, there are MROs in both Al-rich and Co-rich melts due to the noticeable pre-peaks on the plots of structure factors [45]. The structure study of liquid Al-Fe also indicated the Al rich and Fe rich MROs in Al rich melt and Fe rich melt, respectively [49].

On the basis of the aforementioned results in binary liquids, one can assume that there are one or more fictive associates in some melts. The strongest associate locates in the composition (location) with a minimum value of enthalpy of mixing as proposed before [2, 3]. The other associates connect closely to the compounds with high melting points.

## 4 Summary

The characters of molten Ba$_{1-x}$Bi$_x$ with strong ordering trends have been studied in the present work by AIMD simulations at 1123 K. According to structure analyses, e.g., the partial pair correlation functions, it is confirmed that there is a strong chemical affinity between unlike species of Ba and Bi. In addition, the chemical affinity between unlike species in the Ba rich melt is stronger than that in the Bi rich melt. And in the Ba rich melt, the strong chemical affinity induces the Bi atoms barely bond to the same species, resulting in the exclusive formation of the Ba-Bi bonds. The Ba-Ba and Ba-Bi rooted bonded-pairs with large H-A indices (e.g., 1661, 1651, 1551, 1541, 1441, and 1431) are dominant in the Ba rich melt. The bonded-pairs with small H-A indices (e.g., 1421, 1422, 1321, 1311, 1301, 1201, and 1101) are popular in the Bi-rich melt.

In the Ba rich melt, the popular Bi-centered coordination polyhedrons are stable and relatively long-lived with strongly associated to the compounds of $Ba_5Bi_3$ and $Ba_4Bi_3$. The Bi-centered coordination polyhedrons in the Ba rich melt prefer connecting each other with vertex, edge, face, and/or bipyramid sharing. According to bond angle analyses, they are in medium range order. The Ba-centered coordination polyhedrons in the Bi-rich melt also prefer connecting each other, and they are also ordered in medium range, though they are structurally diverse, short-lived, and poorly correlated to $BaBi_3$ compound. These results suggest that it is reasonable to assign two associates, i.e., $Ba_4Bi_3$ and $BaBi_3$ in the Ba-Bi liquid. The former is strongly ordering in medium range and correlated to the compound $Ba_4Bi_3$, the latter is weakly ordering and poorly associated to compound $BaBi_3$.

**Acknowledgements**

The authors acknowledge the financial support by the U.S. Department of Energy (DOE) via Award No. DE-NE0008757, and the National Natural Science Foundation of China (NSFC) via Grant No. 51774202. First-principles calculations were carried out partially on the ACI clusters at the Pennsylvania State University, partially on the resources of the NERSC (National Energy Research Scientific Computing Center) supported by the Office of Science of the U.S. DOE under Contract No. DE-AC02-05CH11231, and partially on the resources of XSEDE (Extreme Science and Engineering Discovery Environment) supported by NSF via Grant No. ACI-1548562.

# Reference

[1] T. Lichtenstein, N.D. Smith, J. Gesualdi, K. Kumar, H. Kim, Thermodynamic properties of Barium-Bismuth alloys determined by emf measurements, Electrochim. Acta, 228 (2017) 628-635.

[2] A.B. Bhatia, W.H. Hargrove, Concentration fluctuations and thermodynamic properties of some compound forming binary molten systems, Phys. Rev. B, 10 (1974) 3186.

[3] C. Bergman, R. Castanet, H. Said, M. Gilbert, J.C. Mathieu, Configurational entropy and the regular associated model for compound-forming binary systems in the liquid state, J. Less Common Metals, 85 (1982) 121-135.

[4] R. Schmid, Y.A. Chang, A thermodynamic study on an associated solution model for liquid alloys, Calphad, 9 (1985) 363-382.

[5] A.D. Pelton, S.A. Degterov, G. Eriksson, C. Robelin, Y. Dessureault, The modified quasichemical model I-Binary solutions, Metall. Mater. Trans. B, 31 (2000) 651-659.

[6] K.P. Kotchi, M. Gilbert, R. Castanet, Thermodynamic behaviour of the Sn-Te, Pb-Te, Sn-Se and Pb-Se melts according to the associated model, J. Less Common Metals, 143 (1988) L1-L6.

[7] A. Bouhajib, A. Nadiri, A. Yacoubi, R. Castanet, Investigation of the short-range order in the Ca–Sb melts, J. Alloys Compd., 287 (1999) 167-169.

[8] K. Ozturk, L.Q. Chen, Z.K. Liu, Thermodynamic assessment of the Al–Ca binary system using random solution and associate models, J. Alloys Compd., 340 (2002) 199-206.

[9] Y. Chen, Y. Liu, M. Chu, L. Wang, Phase diagrams and thermodynamic descriptions for the Bi–Se and Zn–Se binary systems, J. Alloys Compd., 617 (2014) 423-428.

[10] J.M. Liu, P.W. Guan, C.N. Marker, N.D. Smith, N. Orabona, S.L. Shang, H. Kim, Z.K. Liu, Thermodynamic properties and phase stability of the Ba-Bi system: A combined computational and experimental study, J. Alloys Compd., 771 (2019) 281-289.

[11] J.J. Lee, B.J. Kim, W.S. Min, Calorimetric investigations of liquid Cu-Sb, Cu-Sn and Cu-Sn-Sb alloys, J. Alloys Compd., 202 (1993) 237-242.

[12] Y.-B. Kang, A.D. Pelton, Modeling short-range ordering in liquids: The Mg–Al–Sn system, Calphad, 34 (2010) 180-188.

[13] J. Ma, Y. Dai, W. Zhou, J. Zhang, J. Wang, B. Sun, Short range orders in molten Al: An *ab initio* molecular dynamics study, Comput. Mater. Sci., 93 (2014) 97-103.

[14] N. Jakse, A. Pasturel, Ab initio molecular dynamics simulations of local structure of supercooled Ni, J. Chem. Phys., 120 (2004) 6124-6127.

[15] N. Jakse, A. Pasturel, Local order of liquid and supercooled zirconium by *ab initio* molecular dynamics, Phys. Rev. Lett., 91 (2003) 195501.

[16] H.W. Sheng, W.K. Luo, F.M. Alamgir, J.M. Bai, E. Ma, Atomic packing and short-to-medium-range order in metallic gallsses, Nature, 439 (2006) 419-425.

[17] K.N. Lad, N. Jakse, A. Pasturel, Signatures of fragile-to-strong transition in a binary metallic glass-forming liquid, J. Chem. Phys., 136 (2012) 104509.

[18] J. Ma, S. Chen, Y. Dai, J. Zhang, J. Yang, J. Wang, B. Sun, The local structure of molten $Ni_{1-x}Al_x$: An *ab initio* molecular dynamics study, J. Non-Cryst. Solids., 425 (2015) 11-19.

[19] J. Kang, J. Zhu, S.H. Wei, E. Schwegler, Y.H. Kim, Persistent Medium-Range Order and Anomalous Liquid Properties of $Al_{1-x}Cu_x$ Alloys, Phys. Rev. Lett., 108 (2012) 115901.

[20] M. Guerdane, H. Teichler, B. Nestler, Local Atomic Order in the Melt and Solid-Liquid Interface Effect on the Growth Kinetics in a Metallic Alloy Model, Phys. Rev. Lett., 110 (2013) 086105.
43

[21] M. Guerdane, F. Wendler, D. Danilov, H. Teichler, B. Nestler, Crystal growth and melting in NiZr alloy: Linking phase-field modeling to molecular dynamics simulations, Phys. Rev. B, 81 (2010) 224108.

[22] I. Kaban, P. Jóvári, V. Kokotin, O. Shuleshova, B. Beuneu, K. Saksl, N. Mattern, J. Eckert, A.L. Greer, Local atomic arrangements and their topology in Ni–Zr and Cu–Zr glassy and crystalline alloys, Acta Materialia, 61 (2013) 2509-2520.

[23] J. Ma, Y. Dai, J. Zhang, S. Chen, J. Yang, H. Xing, Q. Dong, Y. Han, B. Sun, On the chemical effects in molten $Ni_{1-x}M_x$ alloy, Comput. Mater. Sci., 146 (2018) 158-175.

[24] N. Saunders, A.P. Miodownik, CALPHAD (calculation of phase diagrams): a comprehensive guide, Elsevier, 1998.

[25] M. Born, K. Huang, Dynamical Theory of Crystal Lattices, Oxford University Press, Oxford, 1954.

[26] G. Kresse, D. Joubert, From ultrasoft pseudopotentials to the projector augmented-wave method, Phys. Rev. B, 59 (1999) 1758.

[27] G. Kresse, J. Furthmüller, Efficiency of *ab-initio* total energy calculations for metals and semiconductors using a plane-wave basis set, Comput. Mater. Sci., 6 (1996) 15-50.

[28] G. Kresse, J. Furthmüller, Efficient iterative schemes for *ab initio* total-energy calculations using a plane-wave basis set, Phys. Rev. B, 54 (1996) 11169.

[29] Y. Wang, J.P. Perdew, Correlation hole of the spin-polarized electron gas, with exact small-wave-vector and high-density scaling, Phys. Rev. B, 44 (1991) 13298.

[30] J.P. Perdew, Y. Wang, Accurate and simple analytic representation of the electron-gas correlation energy, Phys. Rev. B, 45 (1992) 13244.

[31] S.J. Clark, M.D. Segall, C.J. Pickard, P.J. Hasnip, M.I. Probert, K. Refson, M.C. Payne, First principles methods using CASTEP, Z. Krist.-Cryst. Mater., 220 (2005) 567-570.

[32] M.P. Allen, D.J. Tildesley, Computer Simulation of Liquid, Oxford University Press, New York, 1989.

[33] X.W. Zou, Z.Z. Jin, Y.J. Shang, Static Structure Factor of Non-Simple Liquid Metals Bi, Ga, Sb, and Sn, phys. stat. sol. (b), 139 (1987) 365-370.

[34] Y. Greenberg, E. Yahel, E. Caspi, C. Benmore, B. Beuneu, M.P. Dariel, G. Makov, Evidence for a temperature-driven structural transformation in liquid bismuth, EPL (Europhysics Letters), 86 (2009) 36004.

[35] Y. Waseda, K. Yokoyama, K. Suzuki, Structure of molten Mg, Ca, Sr, and Ba by X-ray diffraction, Z. Naturforsch. A, 30 (1975) 801-805.

[36] J. Akola, N. Atodiresei, J. Kalikka, J. Larrucea, R.O. Jones, Structure and dynamics in liquid bismuth and $Bi_n$ clusters: A density functional study, J. Chem. Phys., 141 (2014) 194503.

[37] J. Souto, M.M.G. Alemany, L.J. Gallego, L.E. González, D.J. González, Ab initio molecular dynamics study of the static, dynamic, and electronic properties of liquid Bi near melting using real-space pseudopotentials, Phys. Rev. B, 81 (2010) 134201.

[38] N.A. Mauro, V. Wessels, J.C. Bendert, S. Klein, A.K. Gangopadhyay, M.J. Kramer, S.G. Hao, G.E. Rustan, A. Kreyssig, A.I. Goldman, K.F. Kelton, Short- and medium-range order in $Zr_{80}Pt_{20}$ liquids, Phys. Rev. B, 83 (2011) 184109.

[39] N.C. El'ad, Y. Greenberg, E. Yahel, B. Beuneu, G. Makov, What is the structure of liquid Bismuth?, in: J. Physics: Conf. Series, IOP Publishing, 2012, pp. 012079.

[40] J.D. Honeycutt, H.C. Andersen, Molecular dynamics study of melting and freezing of small

Lennard-Jones clusters. , J. Phys. Chem., 91 (1987) 4950-4963.

[41] A.S. Clarke, H. Jónsson, Structural changes accompanying densification of random hard-sphere packings, Phys. Rev. E, 47 (1993) 3975-3984.

[42] J.L. Finney, Random packings and the structure of simple liquids. I. The geometry of random close packing, in:    Proc. R. Soc. Lond. A The Royal Society, 1970, pp. 479-493.

[43] S.K. Das, J. Horbach, M. M.Koza, S.M. Chatoth, A. Meyer, Influence of chemical short-range order on atomic diffusion in Al-Ni melts, Appl. Phys. Lett., 86 (2005) 011918.

[44] Y.Q. Cheng, E. Ma, Atomic-level structure and structure–property relationship in metallic glasses, Prog. Mater. Sci, 56 (2011) 379-473.

[45] O. S. Roik, O.V. Samsonnikov, V.P. Kazimirov, V.E. Sokolskii, Short and medium-range order in liquid binary Al-Ni and Al-Co alloys, J. Mol. Liq. , 145 (2009) 129.

[46] M. Maret, T. Pomme, A. Pasturel, Structure of liquid $Al_{80}Ni_{20}$ alloy, Phys. Rev. B, 42 (1990) 1598.

[47] J. Qin, X. Bian, S.I. Sliusarenko, W. Wang, Pre-peak in the structure factor of liquid Al-Fe alloy, J. Phys.: Conden. Matter, 10 (1998) 1211.

[48] J. Yang, J. Zhang, Y. Dai, J. Ma, F. Li, F. Bian, J. Mi, B. Sun, Ab initio simulation: The correlation between the local melt structure and segregation behavior of Fe, V, Ti and Si in liquid Al, Comput. Mater. Sci., 109 (2015) 41-48.

[49] X. Li, J. Wang, J. Qin, S. Pan, B. Dong, The medium-range orders transition in liquid Fe–Al alloys, Comput. Mater. Sci., 161 (2019) 199-208.

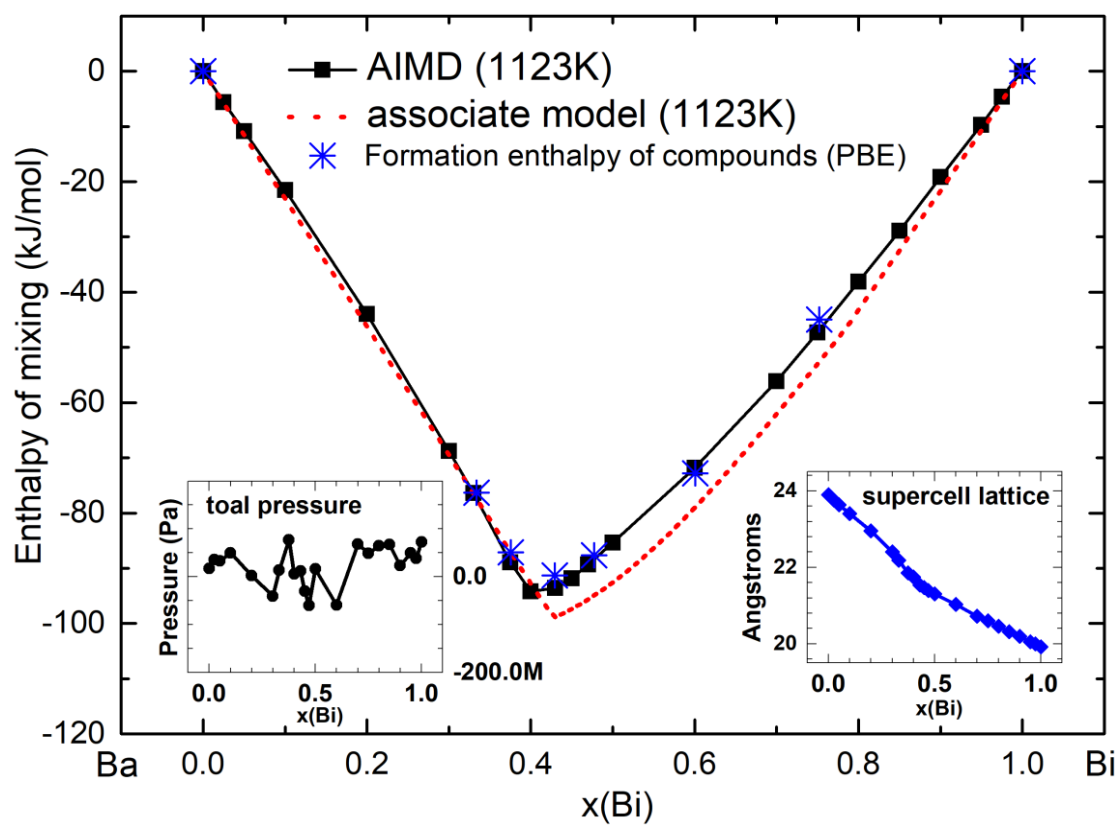

**Figure 1.** Calculated enthalpy of mixing of liquid Ba$_{1-x}$Bi$_x$ by AIMD (solid square, this work) and associate model (dotted-dashed line, CALPHAD modeling [10]) at 1123 K. The star symbols are enthalpy of formation of intermetallic compounds by first-principles calculations using the PBE potential at 300 K [10]. The insets are calculated total pressures (left) and supercell size (right) by AIMD simulations.

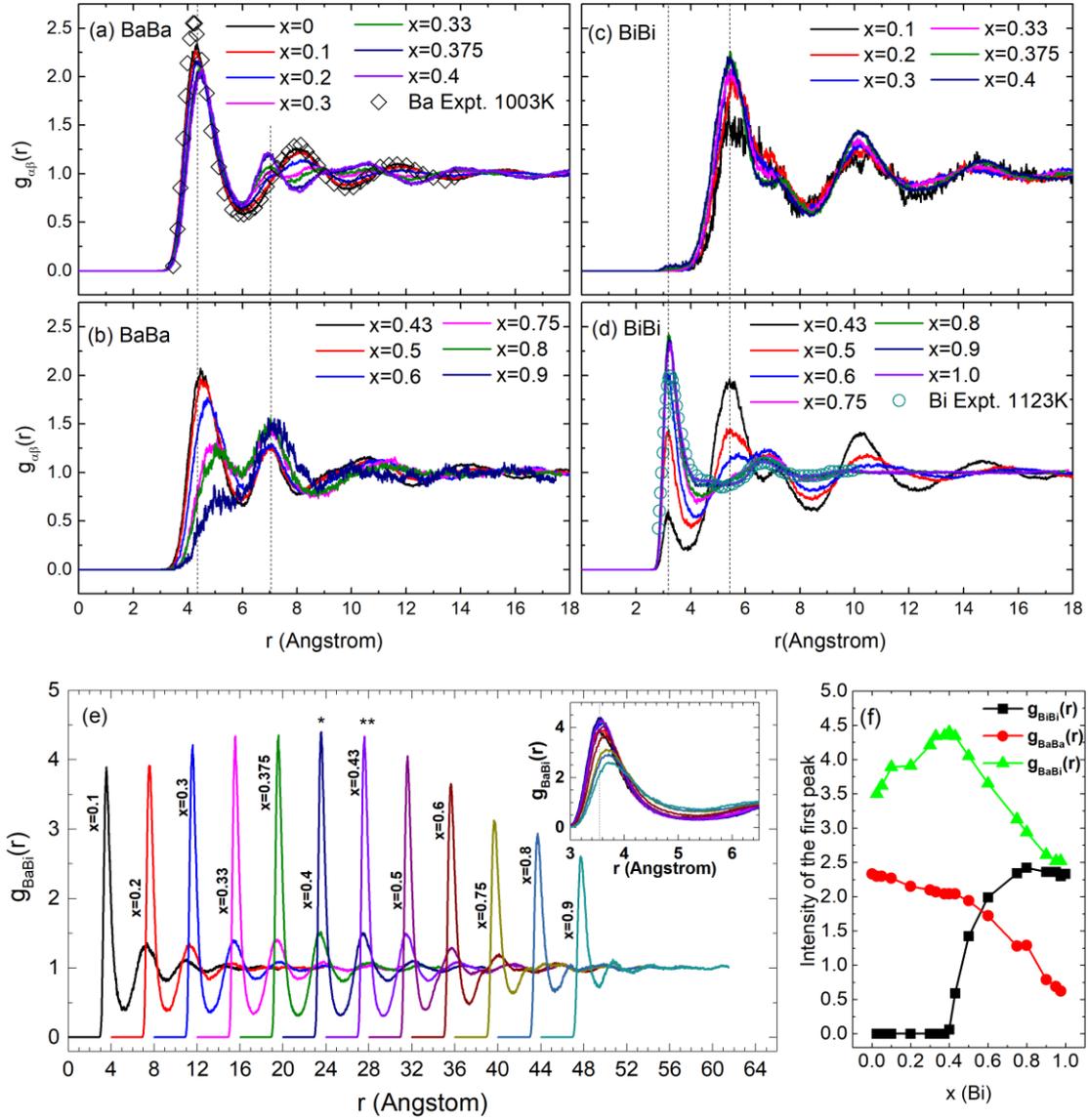

**Figure 2.** Partial PCFs of molten $Ba_{1-x}Bi_x$ at 1123 K, including partial PCFs of $g_{BaBa}(r)$ (a and b), $g_{BiBi}(r)$ (c and d), $g_{BaBi}(r)$ (e), and the comparison of intensities for the first peaks of PCFs (f). The curves of $g_{BaBi}(r)$ in (e) are shifted to right by 4 Å for each curve for the purpose of visualization. The open diamonds in (a) and the open circles in (d) are experimental results of molten Ba by X-ray diffraction [35] and of molten Bi by neutron scattering [34], respectively.

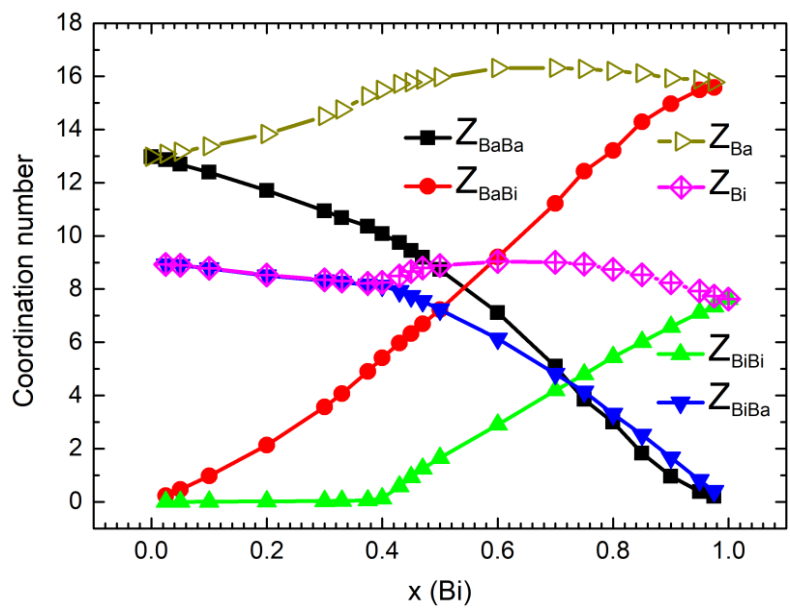

**Figure 3.** Calculated partial coordination numbers $Z_{BaBa}$ (solid square), $Z_{BaBi}$ (solid circle), $Z_{BiBi}$ (solid up triangle), $Z_{BiBa}$ (solid down triangle), $Z_{Ba}$ (open right triangle), and $Z_{Bi}$ (diamond with cross) for molten $Ba_{1-x}Bi_x$ at 1123 K.

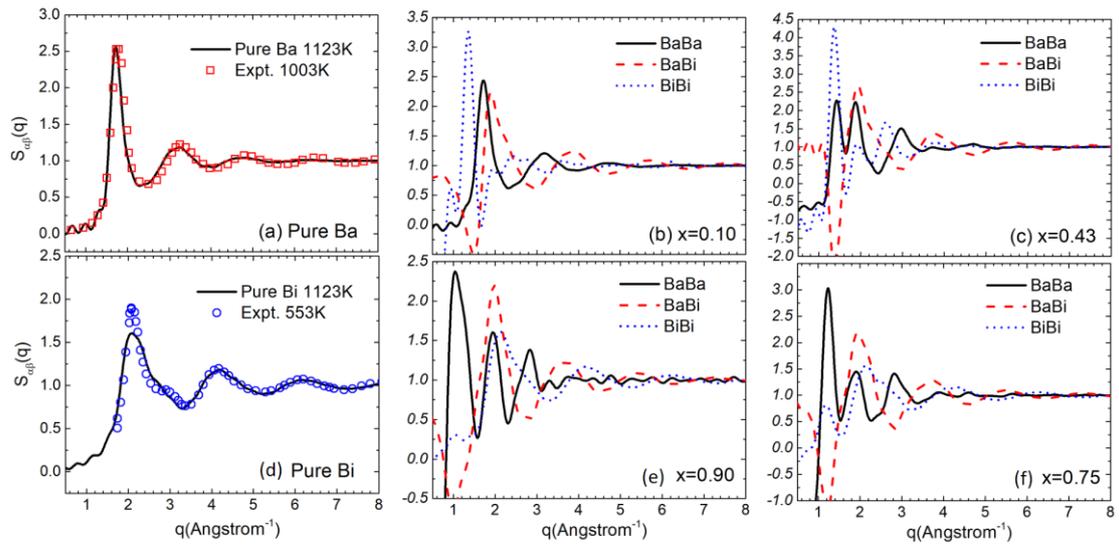

**Figure 4.** Static structure factors of pure Ba (a) and Bi (b), and the partial static structure factors for molten Ba$_{1-x}$Bi$_x$ with $x$ = 0.10 (b), 0.43 (c), 0.75 (e), and 0.90 (f) at 1123 K. The open squares in (a) and the open circles in (d) are experimental result for molten Ba at 1003 K by X-ray diffraction [35] and molten Bi at 553 K by neutron scattering [39], respectively.

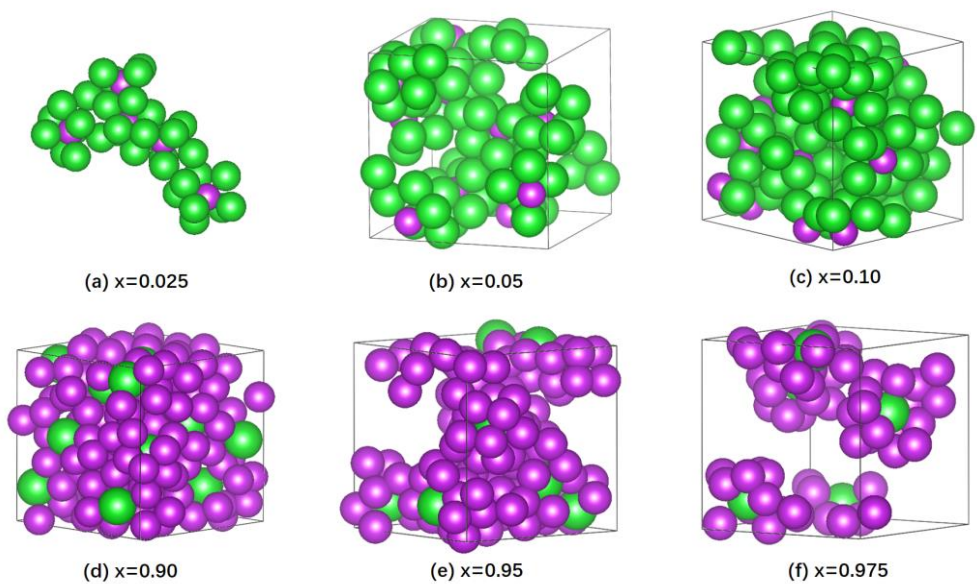

**Figure 5.** Snapshots of AIMD configurations for some connected Bi- and Ba-centered polyhedrons in molten Ba$_{1-x}$Bi$_x$ with $x$ = 0.025 (a), 0.05 (b), 0.10 (c), 0.90 (d), 0.95 (e), and 0.975 (f). The unconnected atoms are removed for clear view; and the green and purple spheres represent Ba and Bi atoms, respectively.

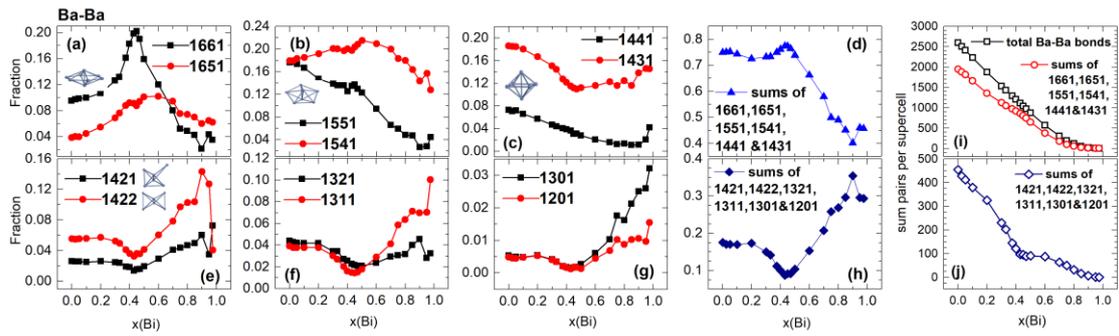

**Figure 6.** Calculated fractions of Ba-Ba bonded-pairs with different H-A indices (a-c and e-g); together with the sum fraction over the pairs with large H-A indices of 1661, 1651, 1551, 1541, 1441, and 1431 (d); the sum fraction of those pairs with small H-A indices of 1421, 1422, 1321, 1311, 1301, and 1201 (h); and the total numbers of bonded-pairs with large H-A indices (i) and small H-A indices (j). The inset motifs in (a), (b), (c), and (e) are for the bonded-pairs with indices 1661, 1551, 1441, 1421 (up one), and 1422 (down one), respectively.

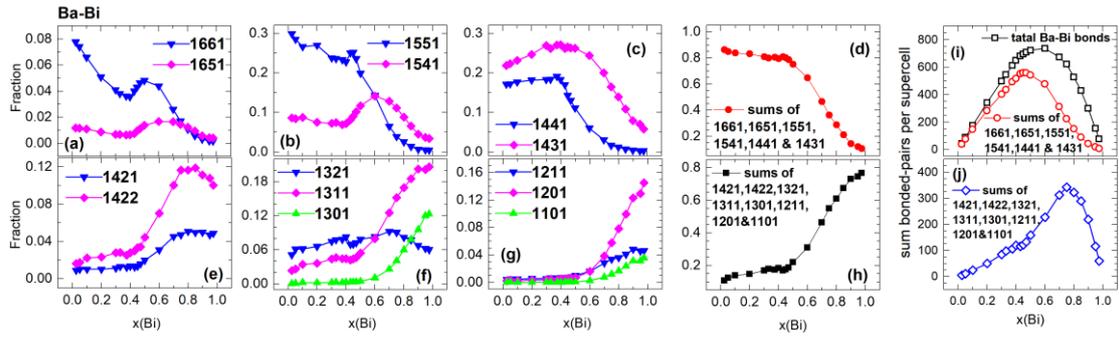

**Figure 7.** Calculated fractions of Ba-Bi bonded-pairs with different H-A indices (a-c and e-g) together with the sum fraction of those pairs with large H-A indices of 1661, 1651, 1551, 1541, 1441, and 1431 (d); the sum fraction of those pairs with small H-A indices of 1421, 1422, 1321, 1311, 1301 1211, 1201, and 1101 (h); and the total numbers of bonded-pairs with large H-A indices (i) and small H-A indices (j).

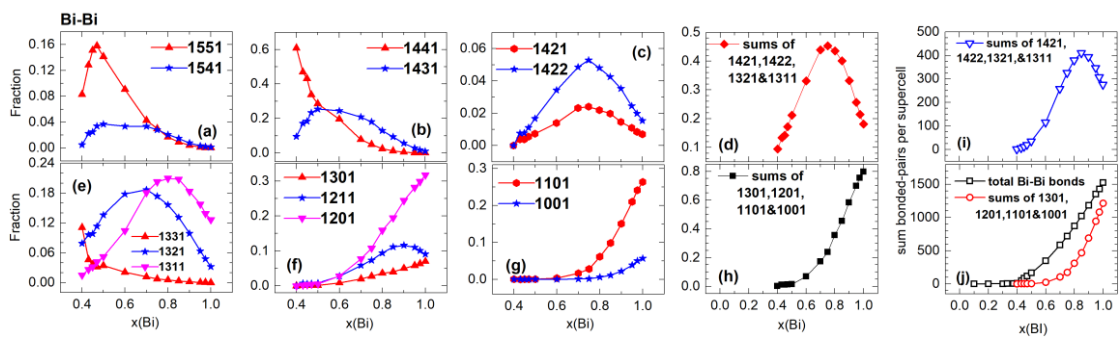

**Figure 8.** Calculated fractions of Bi-Bi bonded-pairs with different H-A indices (a-c and e-g) together with the combined fraction of those pairs with H-A indices of 1421, 1422, 1321, and 1311 (d); the combined fraction of those pairs with small H-A indices 1301, 1201, 1101, and 1001 (h); the sum of bonded-pairs with H-A indices of 1421, 1422, 1321, and 1311 (open down triangles) (i), and 1301,1201,1101, and 1101 (open circles) (j).

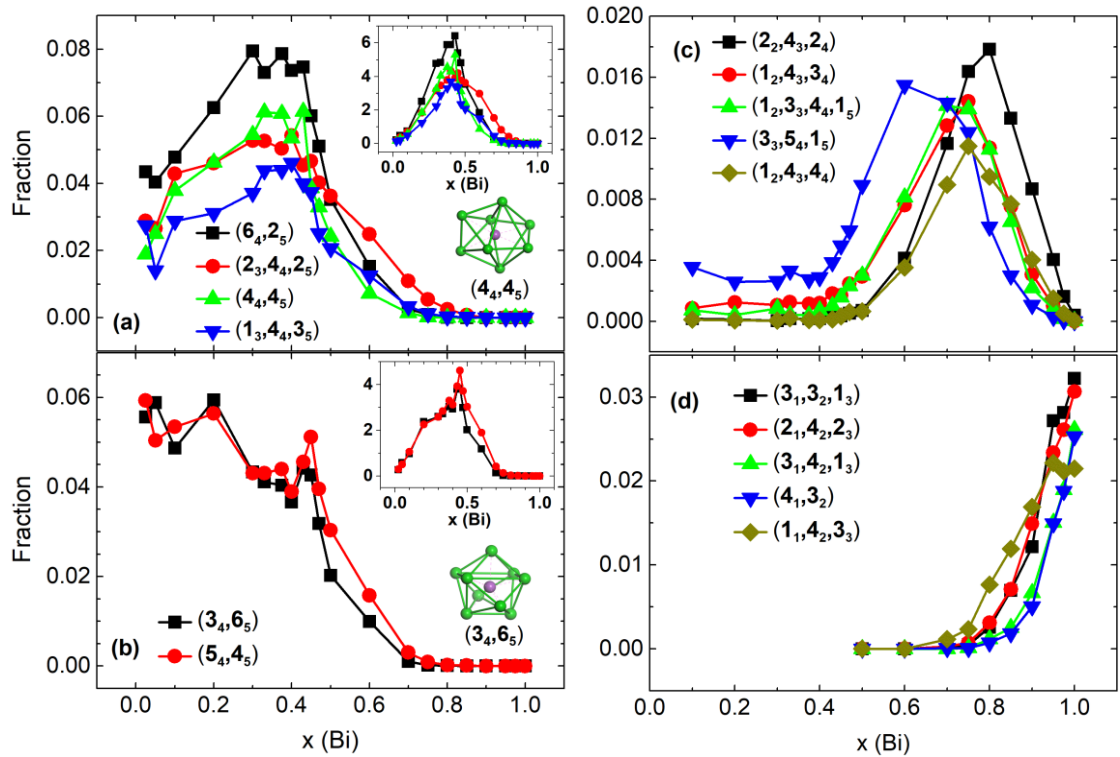

**Figure 9.** Calculated fractions of BIPs including the top six polyhedrons in molten $Ba_4Bi_3$ (a and b), the top five ones in molten $BaBi_3$ (c and d). The insets in (a) and (b) are the numbers for the corresponding polyhedra per supercell. The motifs of $(4_4,5_4)$ and $(3_4,6_4)$ are shown in (a and b).

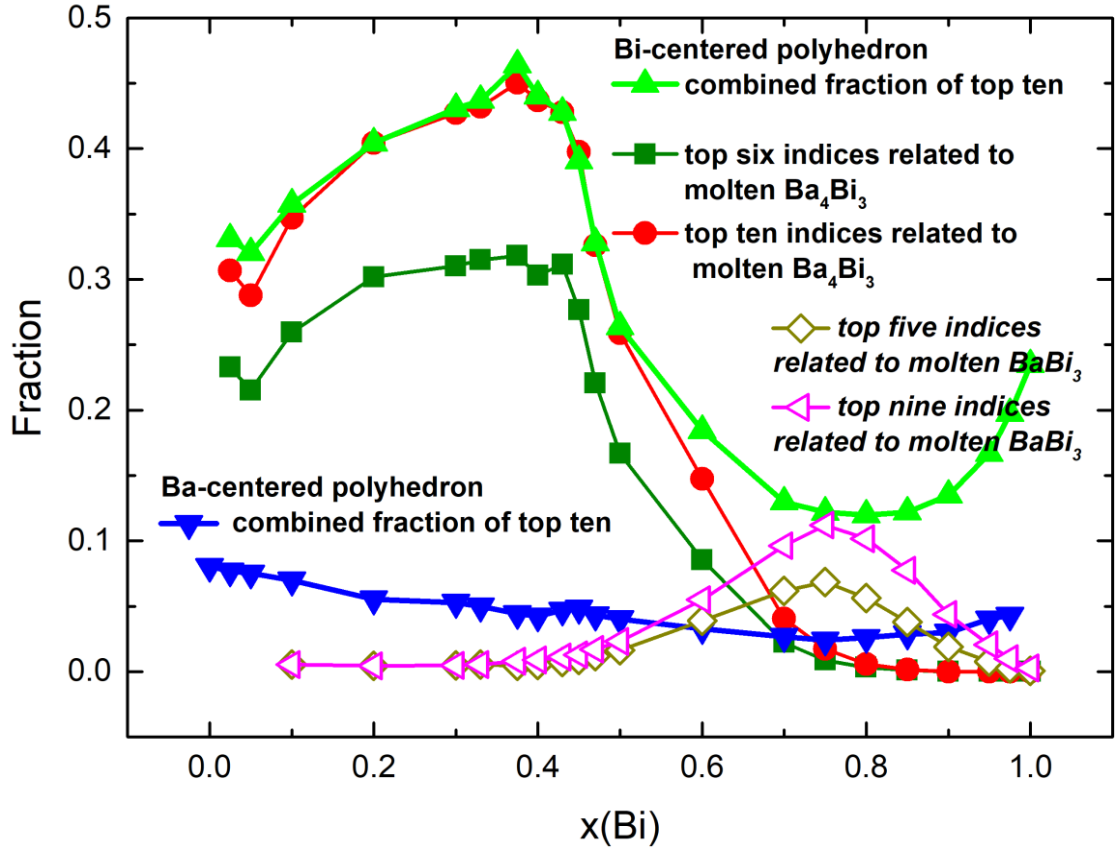

**Figure 10.** Combined fraction of the dominant BIPs and BAPs with a function of the mole fraction of Bi (*x*). The filled up triangles are for the top ten BIPs; the filled down triangles are for the top ten BAPs; the filled circles and filled squares are for BIPs with the top ten and the top six indices related to molten $Ba_4Bi_3$, respectively; and the open left triangles and the open diamonds are for BIPs with the top nine and the top five related to molten $BaBi_3$, respectively.

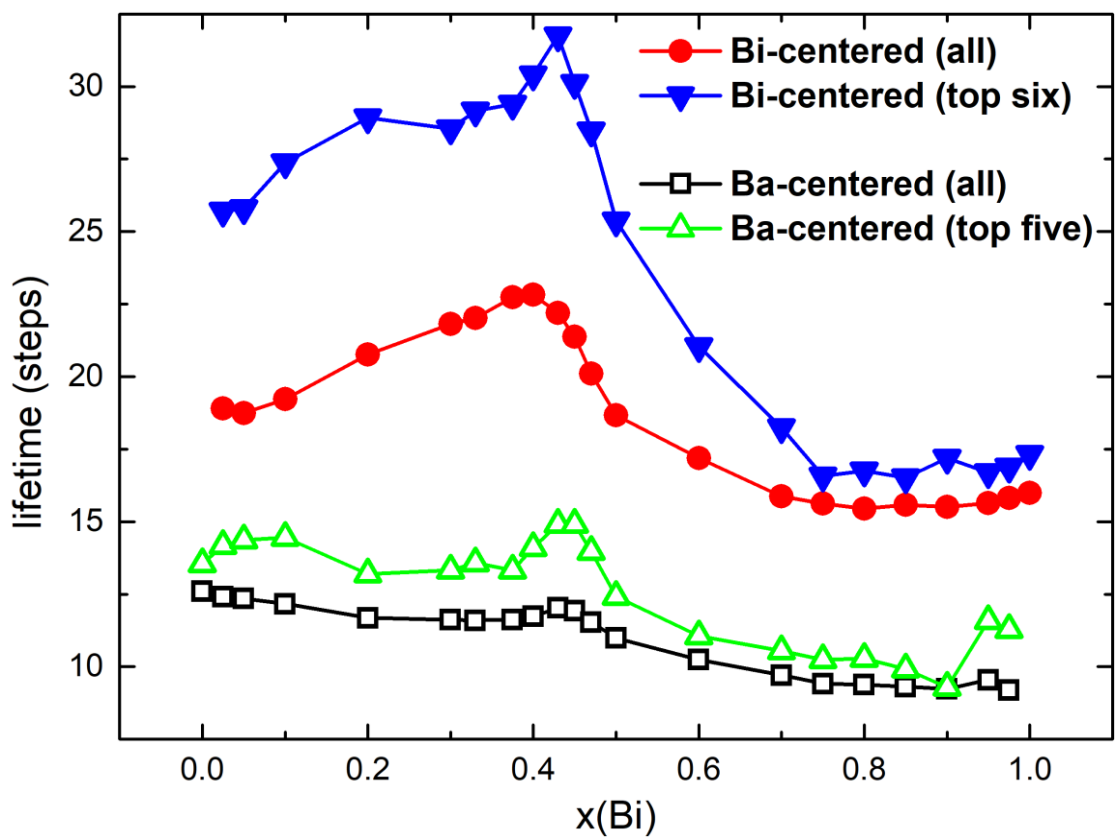

**Figure 11.** Average lifetimes of Bi- and Ba-centered polyhedrons as a function of Bi concentration, where the filled circles represent the average over all Bi-centered polyhedrons; the filled down triangles are the average over the top six Bi-centered polyhedrons; the open squares are the average over all Ba-centered polyhedrons; and the open up triangles are the average over the top five Ba-centered polyhedrons.

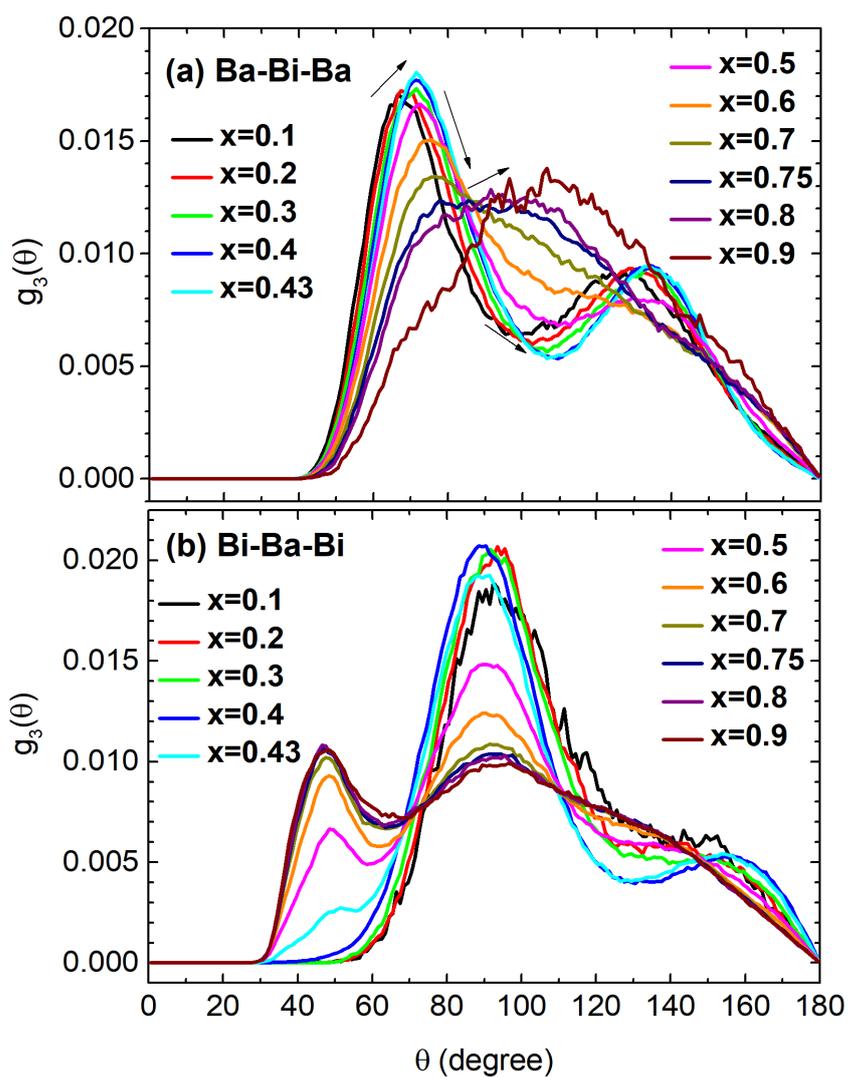

**Figure 12.** Comparison of the bond angle distribution functions for the triples of Ba-Bi-Ba (a), and Bi-Ba-Bi (b).

**Table 1.**

Indices and fraction (*f*) of bonded-pairs and the coordination polyhedrons for crystalline Ba (BCC, Im$\bar{3}$m), Bi (Rhom_A7, R$\bar{3}$m), Ba$_2$Bi, Ba$_5$Bi$_3$, Ba$_4$Bi$_3$, Ba$_{10}$Bi$_{11}$, and BaBi$_3$.

| Mater. | Space Group | Common neighbor analysis ||||||  Coordination polyhedron ||||
|       |       | Ba-Ba || Ba-Bi || Bi-Bi || Ba-centered || Bi-centered ||
|       |       | Index | *f* | Index | *f* | Index | *f* | Index | *f* | Index | *f* |
| Ba | Im$\bar{3}$m | 1661 | 0.571 | | | | | $(6_4,8_6)$ | 1.0 | | |
|    |              | 1441 | 0.429 | | | | | | | | |
| Ba$_2$Bi | I4/mmm | 1661 | 0.50  | 1551 | 0.889 | | | $(1_4,8_5,8_6)$ | 0.5 | $(1_4,8_5)$ | 1.0 |
|    |              | 1551 | 0.167 | 1441 | 0.111 | | | $(4_4,4_5,4_6,4_8)$ | 0.5 | | |
|    |              | 1441 | 0.167 | | | | | | | | |
|    |              | 18xx | 0.166 | | | | | | | | |
| Ba$_5$Bi$_3$ | P6$_3$/mcm | 1661 | 0.538 | 1551 | 0.667 | | | $(5_4,6_5,4_6)$ | 0.6 | $(3_4,6_5)$ | 1.0 |
|    |              | 1541 | 0.231 | 1441 | 0.333 | | | $(3_4,6_5,8_6)$ | 0.4 | | |
|    |              | 1441 | 0.115 | | | | | | | | |
|    |              | 1422 | 0.115 | | | | | | | | |
| Ba$_4$Bi$_3$ | I$\bar{4}$3d | 1771 | 0.273 | 1551 | 0.5 | | | $(3_4,9_5,2_6,3_7)$ | 1.0 | $(4_4,4_5)$ | 1.0 |
|    |              | 1661 | 0.182 | 1441 | 0.5 | | | | | | |
|    |              | 1551 | 0.545 | | | | | | | | |
| Ba$_{11}$Ba$_{10}$ | I4/mmm | 1661 | 0.52 | 1661 | 0.0253 | 1441 | 0.778 | $(1_4,10_5,6_6)$ | 0.3636 | $(2_4,8_5)$ | 0.4 |
|    |              | 1551 | 0.44 | 1551 | 0.7595 | 1431 | 0.222 | $(1_4,10_5,4_6)$ | 0.3636 | $(3_4,6_5)$ | 0.2 |
|    |              | 1541 | 0.04 | 1541 | 0.0506 | | | $(2_4,10_5,5_6)$ | 0.1818 | $(4_4,6_5,1_6)$ | 0.2 |
|    |              | | | 1441 | 0.1139 | | | $(1_4,12_5,4_6)$ | 0.0909 | $(5_4,4_5)$ | 0.1 |
|    |              | | | 1431 | 0.0506 | | | | | $(4_4,4_5)$ | 0.1 |
| BaBi$_3$ | P4/mmm | 1441 | 1.0 | 1661 | 1.0 | 1431 | 1.0 | $(6_4,12_6)$ | 1.0 | $(8_4,4_6)$ | 1.0 |
| Bi | R$\bar{3}$m | 1001 | 1.0 | | | | | | | $(6_0)$ | 1.0 |